%% file: main.tex
\documentclass[twoside,11pt]{article}

\usepackage[preprint]{jmlr2e} 

\usepackage{amsmath}
\usepackage{graphicx,psfrag,epsf}
\usepackage{enumerate}
\usepackage{natbib}
\usepackage{url} 
\usepackage{bbm}
\usepackage{subcaption} 
\usepackage[linesnumbered,ruled]{algorithm2e}
\usepackage{rotating}
\usepackage{booktabs}
\usepackage{xcolor}
\usepackage{hyperref}
\usepackage{float}
\usepackage{verbatim}
\usepackage[normalem]{ulem}
\useunder{\uline}{\ul}{}
\usepackage{dirtytalk}

\newcommand{\Focus}{$\mbox{FOCuS}$} 
 
\newcommand{\FocusZ}{$\mbox{FOCuS}^0$} 
 
\newcommand{\RFocus}{$\mbox{R-FOCuS}$}

\newcommand{\arr}{\xleftarrow{}}

\providecommand{\e}[1]{\ensuremath{\times 10^{#1}}}
\SetKwProg{Init}{init}{}{}

\usepackage{tikz}
\usetikzlibrary{arrows}
\jmlrheading{1}{2021}{999}{15/10}{00/00}{paper00}{Gaetano Romano, Idris Eckley, Paul Fearnhead, Guillem Rigaill}

\ShortHeadings{Fast Functional Pruning for Online Changepoints}{Romano, Eckley, Fearnhead, Rigaill}

\firstpageno{1}

\begin{document}

\title{Fast Online Changepoint Detection via Functional Pruning CUSUM Statistics}

\author{\name Gaetano\ Romano \email g.romano@lancaster.ac.uk\\
        \addr Department of Mathematics and Statistics,\\ Lancaster University,\\ Lancaster, United Kingdom, LA1 4YF
        \AND
        \name Idris\ Eckley \email i.eckley@lancaster.ac.uk\\
        \addr Department of Mathematics and Statistics,\\ Lancaster University,\\ Lancaster, United Kingdom, LA1 4YF
        \AND
        \name Paul\ Fearnhead \email p.fearnhead@lancaster.ac.uk\\
       \addr Department of Mathematics and Statistics,\\ Lancaster University,\\ Lancaster, United Kingdom, LA1 4YF
       \AND
       \name Guillem\ Rigaill \email guillem.rigaill@inrae.fr \\
       \addr Université Paris-Saclay, CNRS, INRAE, Univ Evry, \\ Institute of Plant Sciences Paris-Saclay (IPS2), \\ 91405, Orsay, France}

\editor{Editor Name}

\maketitle

\begin{abstract}
Many modern applications of online changepoint detection require the ability to process high-frequency observations, sometimes with limited available computational resources. Online algorithms for detecting a change in mean often involve using a moving window, or specifying the expected size of change. Such choices affect which changes the algorithms have most power to detect. We introduce an algorithm, Functional Online CuSUM (FOCuS), which is equivalent to running these earlier methods simultaneously for all sizes of window, or all possible values for the size of change. Our theoretical results give tight bounds on the expected computational cost per iteration of FOCuS, with this being logarithmic in the number of observations. We show how FOCuS can be applied to a number of different change in mean scenarios, and demonstrate its practical utility through its state-of-the art performance at detecting anomalous behaviour in computer server data.
\end{abstract}

\begin{keywords}
Breakpoints; Changepoints; CUSUM; Online; Real-time analysis; FPOP. \end{keywords}


\section{Introduction}

Over the previous decade we have witnessed a renaissance of changepoint algorithms, and they can now be seen to make a difference to many real-world applications. Most of the current literature focuses on an \textit{a posteriori} analysis, having observed a series of data. Such an approach is often referred to as \textit{offline} changepoint detection.
However, as technology develops, the demand from several fields for \textit{online} changepoint detection procedures has increased drastically over recent years. Examples include, but are not limited to, IT and cyber security  \cite[]{jeske2018statistical,tartakovsky2012efficient,peng2004proactively}; 
detecting gamma ray bursts in astronomy \cite[]{fridman2010method,fuschino2019hermes}; detecting earthquake tremors \cite[]{popescu2017new,xie2019asynchronous}; 
industrial processes monitoring \cite[]{pouliezos2013real}; detecting adverse health events  \cite[]{clifford2015physionet};
and monitoring the structural integrity of aeroplanes \cite[]{alvarez2020flight, basseville2007flight}. 

The online setting raises computational challenges that are not present in offline changepoint detection. A procedure needs to be sequential, in the sense that one should process the observations as they become available, and at each iteration one should make a decision whether to flag a changepoint based on the information to date. The procedure also needs to be able to run on a finite state machine for an indefinite amount of iterations, \textit{i.e.} be constant in memory. Finally, a procedure should ideally be able to process observations, at least on average, as quickly as they arrive. Many online changepoint application settings have high frequency observations, and some also have limited computational resources.  For example, the observations from ECG data in the 2015 PhysioNet challenge \cite[]{clifford2015physionet} are sampled at 240Hz, while methods for detecting gamma ray bursts \cite[]{fuschino2019hermes} need to process high-frequency observations and be able to be run on small computers on board micro-satellites. A challenge with online changepoint algorithms is to meet such computational constraints whilst still having close to optimal statistical properties. This paper considers the univariate change in mean problem within precisely this setting.

Current online changepoint methods with a linear computational cost include the method of \cite{page1955test} that assumes knowledge both of the pre-change and post-change mean; or moving window methods such as MOSUM. Assuming the pre-change mean is known is reasonable in many applications as there will be substantial data to estimate this mean \cite[though see discussion in][]{gosmann2019new}. However the method of \cite{page1955test} can lose power if the assumed size of the change is wrong. For example, this method can have almost no power to detect changes that are less than half the size of the assumed change. Similarly, moving window methods can perform poorly if the window size is inappropriate for the size of the change. For example, a small window size will result in little power at detecting small changes, whilst too large a window will lead to delays in detecting larger changes. See Section \ref{sec:background}
for an example of these issues.

An alternative approach, with more robust statistical properties is to, e.g., apply a moving window but consider all possible window sizes. In the known pre-change mean setting this is known as the Page-cusum approach \cite[]{kirch2018modified} and is the approach of \cite{yu2020note} for the case of an unknown pre-change mean. The theoretical results in \cite{yu2020note} demonstrate the excellent statistical properties of such a method. However current exact implementations of this idea have a computational cost per iteration that is linear in the number of observations, and thus have an overall quadratic computational cost.  The problem of deriving an efficient implementation capable of solving the Page-cusum statistic dates back to the past century. \cite{basseville1993detection} state that \say{[...] the GLR [Generalized Likelihood Ratio - Page's method] algorithm is computationally complex}, before presenting some approximations. Recently, \cite{yu2020note} comment on the challenge of developing faster algorithms with good statistical guarantees: \say{we are not aware of nor expect to see any theoretically-justified methods with linear order computational costs}. This paper presents such an algorithm, which we call Functional Online CuSum (\Focus{}). We develop \Focus{} for detecting changes in mean in univariate data  under a Gaussian model, and it can be applied to settings where either the pre-change mean is known or unknown. \Focus{} recursively updates a piecewise quadratic, whose maximum is the test statistic for a change. We show that the amortized cost for solving the recursion is constant per iteration. Maximising the function has a computational cost proportional to the number of components in the piecewise quadratic, and the average number of components increases with the logarithm of the number of observations. We can obtain algorithms with a constant computation per iteration by, when necessary, restricting the maximisation to a fixed number of components. We develop one such approximate version of \Focus{} and show empirically that it has almost identical statistical performance to an exact implementation. In the unknown pre-change mean case \Focus{} implements the method of \cite{yu2020note}, and our exact implementation of \Focus{} can analyse 1 million observations in less than a second on a common personal computer. 

Much research on online changepoint methods has looked at how to implement methods so that they have well characterised performance under the null hypothesis of no change. There are two distinct criteria for quantifying a method's behaviour under the null, one is the average run length \cite[]{reynolds1975approximations} which is the expected number of observations until we detect a change. The other is a significance level -- the probability of ever detecting a change if the method is run on infinitely long data. In practice these two criteria affect the choice of threshold for a detection method. If we wish to control the significance level then we need a threshold that increases with the number of observations \cite[see e.g.][]{kirch2015use}, whereas if we wish to control the average run length we can use a fixed threshold. The \Focus{} algorithm can be used with either approach -- but for simplicity we will only use a fixed threshold in this paper. 

The outline of the paper is as follows. In Section \ref{sec:pre-change-known} we consider the challenge of detecting a univariate change in mean when the pre-change mean is assumed known. We present the \Focus{} algorithm which can be viewed as implementing the procedure of \cite{page1955test} simultaneously for all possible size of change. Our main theoretical result shows that \Focus{} achieves this with an average computational cost per iteration that is logarithmic in the number of data points. Our bound on the average per iteration cost is tight, and processing the one-millionth observation roughly equates to the cost of evaluating 15 quadratics. 
 In Section \ref{sec:pre-change-unknown} we study FOCuS for the case where the pre-change mean is unknown. This algorithm can be viewed as implementing the statistical tests of \cite{yu2020note}. Interestingly, whilst their algorithms are either exact but with a linear cost per iteration, or approximate with a cost that is logarithmic in the number of iterations, the \Focus{} algorithm is both exact and has an average cost that is logarithmic per iteration. We do not present any statistical theory for \Focus{}, as this is covered in \cite{yu2020note}.
In Section \ref{sec:focus_extensions} we then introduce an extension of the \Focus{} algorithm to the scenario of detecting changes in the presence of outliers.  In Section \ref{sec:application} we show a monitoring application for \Focus{} on some AWS Cloudwatch server instances.  Finally, the paper concludes with a discussion,  which includes evaluating the use of running \Focus{} independently on multiple data streams and then combining the results to detect a change that may jointly affect multiple streams.

Software implementing FOCuS and the code for our simulation study is available at \url{https://github.com/gtromano/FOCuS}.

\section{Known pre-change mean}
\label{sec:pre-change-known}

\subsection{Problem Set-up and Background}
\label{sec:background}

Consider the problem of detecting a change in mean in univariate data. We will let $x_t$ denote the data at time $t$, for $t=1,2,\ldots$. We are interested in online detection, that is after observing each new data point we wish to decide whether or not to flag that a change has occurred.  We first assume that the pre-change mean is known. Often the methods below are implemented in practice using a plug-in estimator for the pre-change mean that is calculated from training data.

Whilst there are many different approaches to online detection of change \cite[see][for some examples]{veeravalli2014quickest}, a common approach \cite[see][]{kirch2018modified} is to use a cumulative sum of score statistics, also know as a CUSUM based procedure. Assume we model our data as coming from a parametric model with density $f(x;\mu)$ and denote the pre-change mean as $\mu_0$. Define the score statistic of an observation $x$ as 
\[
H(x,\mu)=\frac{\partial \log f(x;\mu)}{\partial \mu}.
\]
Then if there is no change prior to time $n$ 
\[
\mbox{E}\left( H(X_i,\mu_0) \right)=0,\mbox{ for $i=1,\ldots,n$.}
\]
Thus evidence of a change prior to time $n$ can be obtained by monitoring the absolute values of partial sums of these score statistics, which we denote as
\[
S(s,n)=\sum_{i=s+1}^n H(x_i,\mu_0).
\]
The idea is that these partial sums should be close to 0 if there is no change, and conversely diverge from zero if there is a change.

For ease of presentation, and to make ideas concrete, in the following we will consider the case where we have a Gaussian model { with unit variance} for the data. In this case $H(x,\mu)=(x-\mu)$. Also as we are assuming $\mu_0$ is known, then without loss of generality we can set $\mu_0=0$.

There have been a number of different choices of partial sums that we can monitor. For detecting a change after observing $x_n$, \cite{kirch2018modified} highlight the following statistics:
\begin{eqnarray}
    \text{CUSUM} & \quad & C(n)=\frac{1}{\sqrt{n}}|S(0, n)|; \label{eq:CUSUM} \\
\text{MOSUM} & & M_w(n)=\frac{1}{\sqrt{w}}|S(n - w, n)|; \label{eq:MOSUM}\\
\text{mMOSUM} & & M^{(\text{m})}_k(n)=\frac{1}{\sqrt{n k}} |S(n-\lfloor kn \rfloor , n)|; \label{eq:mMOSUM}\\
\text{Page-CUSUM} & & P(n)=\max_{0 \leq w < n} \ \frac{1}{\sqrt{w}} | S(n-w, n)|. \label{eq:page-CUSUM}
\end{eqnarray}
The scale factor in each case is to normalise the cumulative sum $S(\cdot,\cdot)$, with the aim of standardizing its variance.
In each case we would compare the statistic at time $n$ with some appropriate threshold, and detect a change prior to $n$ if the statistic is above the threshold. As discussed in the introduction the choice of threshold impacts the properties of the test under the null. It is possible to choose thresholds that are constant or that increase as $n$ increases \cite[]{kirch2018modified}, but for simplicity we will use constant thresholds throughout.

The standard CUSUM statistic uses the partial sum of score statistics to time $n$. For both the MOSUM procedure (\cite{eiauer1978use}, \cite{chu1995mosum}) and mMOSUM procedure \cite[originating in][]{chen2010modified}  we need to specify a tuning parameter. The MOSUM method uses the partial sum over a window of the most recent $w>0$ observations, whilst the mMOSUM fixes some proportion $0<k<1$ and uses the partial sum over the most recent proportion $k$ of the observations. All three of these statistics are online, in that there is only an $O(1)$ update of the statistics as we process each new data point. The Page-CUSUM maximises over all possible partial sums ending at time $n$.

\begin{figure}[!htb]
    \centering
    \begin{subfigure}[b]{\linewidth}
        \centering
        \includegraphics[width=.7\linewidth]{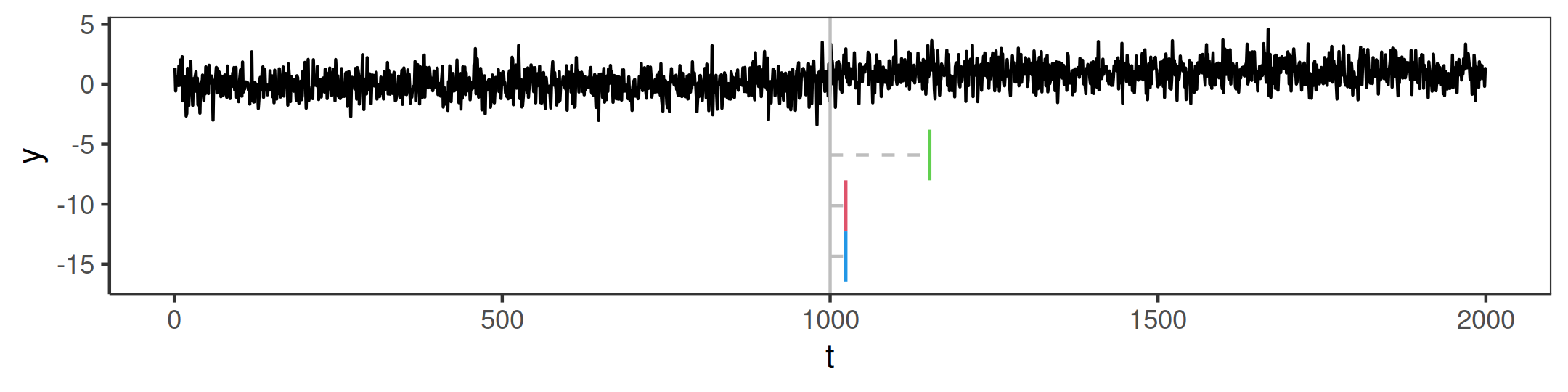}
        \caption{}
        \label{fig:introexe1}
    \end{subfigure}
    \begin{subfigure}[b]{\linewidth}
        \centering
        \includegraphics[width=.7\linewidth]{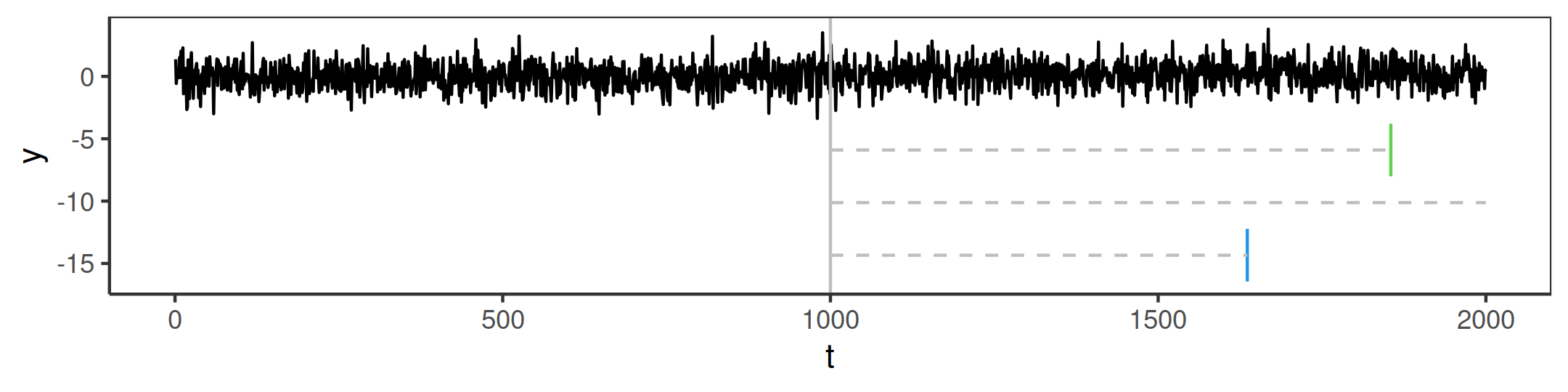}
        \caption{}
        \label{fig:introexe2}

    \end{subfigure}
    \begin{subfigure}[b]{\linewidth}
        \centering
        \includegraphics[width=.7\linewidth]{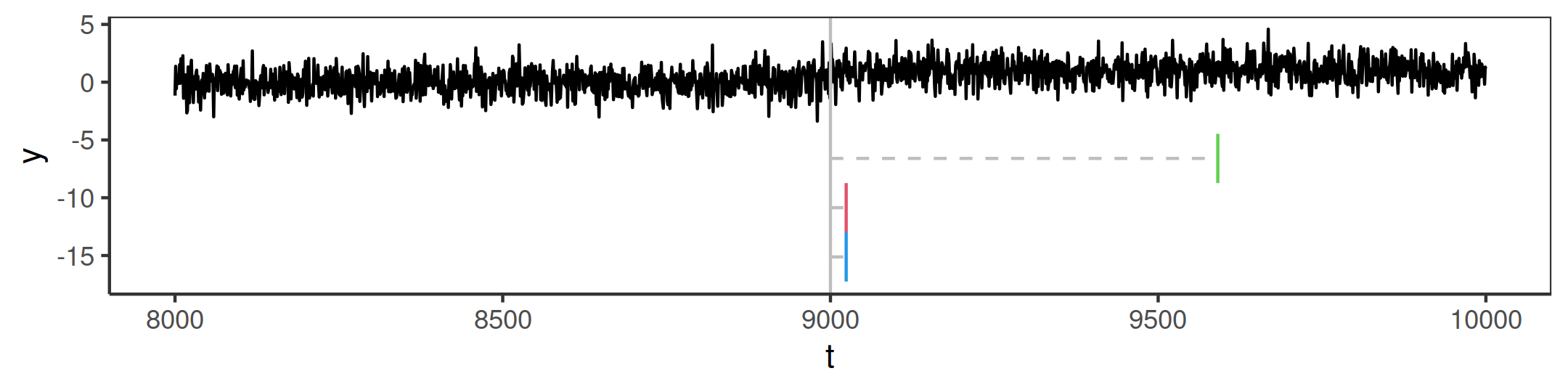}
        \caption{}
        \label{fig:introexe3}

    \end{subfigure}    
    \centering
    \caption{Detection delays of CUSUM (in green), MOSUM (in red, with $w = 50$) and Page-CUSUM (in blue) on three sequences. The sequences were generated in the following way: (a) a sequence of 2000 observations with a change of size 1 at 1000; (b) similar to (a) but with a change in the mean of 0.2; (c) similar (a) again, but with an additional $8\e{3}$ observations at the start of the sequence. Thresholds were tuned accordingly to the simulation study in Section \ref{sec:simulation-study}. The solid grey line refers to the true changepoint location, the dashed segments to the detection delays of the super-mentioned procedures.}
    \label{fig:intro-example}
\end{figure}

To illustrate the difference between CUSUM, MOSUM and Page-CUSUM we implement these methods to detect a change at time $t$ to some mean $\mu_1$ for different values of $t$ and $\mu_1$.  We calculate the different test statistics under our model where the data is independent Gaussian with unit variance. In Figure~\ref{fig:intro-example} we compare the detection delay of each statistic.
In Figure \ref{fig:introexe1}, we simulated data with a change after 1000 observations and the size of change is chosen to give high power for the window size of the MOSUM procedure. MOSUM and Page-CUSUM tend to detect a change quickly. In the second example, shown in Figure \ref{fig:introexe2}, we reduce the magnitude of the change. Here the MOSUM test loses power substantially and { fails to detect a change. Whilst increasing the window would detect the change, using a larger window size would increase detection delay in the first example.} In our final example, Figure \ref{fig:introexe3}, we have the same size of change as the first example, however the change now occurs after 8,000 observations. In this case we see that the CUSUM statistic behaves poorly. This is because the CUSUM statistic has to average the signal from data after the change with all the data prior to change, and this reduces the power of the test statistic. This is particularly the case when there is substantial data prior to the change. Both MOSUM and Page-CUSUM perform as in the first example. Whilst we do not show the performance of mMOSUM in these examples, it shares a similar sensitivity to the choice of window proportion, $k$, as the MOSUM does for window size.

The Page-CUSUM approach tries to avoid the issues with choosing a window size within the MOSUM method, and is equivalent to maximising the MOSUM statistic over $w$. However current implementations of Page-CUSUM are not online. Indeed, the computational cost of calculating $\max_{0 \leq s < n} \ | S(s, n)|$ increases linearly with $n$, resulting in an $O(n^2)$ computational complexity.

An alternative approach  \cite[]{Page,page1955test} to detecting the change is based on sequentially applying a likelihood ratio test under an assumed value for the post-change mean, $\mu_1$. Under our Gaussian model with pre-change mean being 0 and variance 1, the contribution to the log-likelihood ratio statistic from a single data point, $x_t$, is
\begin{equation*}
\label{eq:log-likelihood-ratio}
    LR(x_t, \mu_1) = 2 \mu_1\left(x_t -\frac{\mu_1}{2} \right)
\end{equation*}


As is common in this setting, we will work with a test-statistic that is half the log likelihood-ratio statistic (though obviously using such a test statistic is equivalent to using the log likelihood-ratio test statistic).
At time $n$, our test-statistic for a change at time $s$ is thus half the sum of the $LR(x_t,\mu_1)$ terms from $t=s+1,\ldots,n$. As we do not know the time of the change, we maximise over $s$:
\[
\mathcal{Q}_{n, \mu_1} = \max_{0 \leq s \leq n} \ \sum_{t=s+1}^n \mu_1 \left(x_t - \frac{\mu_1}{2}\right),
\]
where we use the convention that the sum from $s=n+1$ to $n$ is 0.
We will call this statistic the {\em sequential-Page statistic}.

Whilst $\mathcal{Q}_{n,\mu_1}$ involves a sum over $n$ terms, \cite{Page} showed that we can calculate $\mathcal{Q}_{n, \mu_1}$ recursively in constant time as:
\begin{equation}
    \label{eq:seq_page-gaus}
    \mathcal{Q}_{n, \mu_1} = \max \left\{ 0, \ \mathcal{Q}_{n - 1, \mu_1} + \mu_1 \left( x_n - \frac{\mu_1}{2}\right) \right\} .
\end{equation}

\noindent One issue with the sequential-Page statistic is the need to specify $\mu_1$, and a poor choice of $\mu_1$ can substantially reduce the power to detect a change. This is similar to the choice of window size for MOSUM.
To partially overcome this, in both cases we can implement the methods multiple times, for a grid of either window sizes or values of $\mu_1$. Obviously, this comes with an increased computational cost.

{
\cite{Lorden} suggest a procedure for extending Page's method to test for all changes bigger than a minimum size of change. To ease exposition assume that we wish to test for a positive change, with a similar approach being applicable for a negative change. Let $\mu_1$ be the minimum size of change. \cite{Lorden} uses a statistic that is the maximum of the sequential-Page statistics over all changes larger than $\mu_1$. This procedure is based on running Page's method for $\mu_1$, and recording the reset times, that is the times at which the sequential-Page statistic is 0. \cite{Lorden} then uses the property that at any time $t$, the maximum of sequential-Page statistic will be the maximum of half the log likelihood-ratio statistic for a change at time $s$, where we maximise over all $s$ greater than or equal to the most recent reset time. The computational cost of this at time $t$ is proportional to the number of time-points since the last reset. 

This idea has some similarity to the \FocusZ{} algorithm we introduce next, in that it tries to cleverly choose which tests to perform. But \FocusZ{} does not require a minimum size of change to be specified; and \FocusZ{} is more efficient at deciding which tests to calculate, so that it is computationally more efficient (unless the minimum size of change of Lorden's algorithm is around 0.5 or higher, see Appendix \ref{App:Lorden}).
}
\subsection{\FocusZ{}: solving the Page recursion for all $\mu_1$} \label{sec:focus0_computational}

Our idea is to solve the sequential-Page recursion simultaneously for all values of the post-change mean. To this end we re-write (\ref{eq:seq_page-gaus}) in terms of a recursion for a function $Q_n(\mu)$ of the post-change mean $\mu_1=\mu$. We then have $Q_0(\mu)=0$ and for $n=1,\ldots,$
\begin{equation}
        \label{eq:functional-recursion}
        Q_n(\mu) = \max \left\{ 0, \ Q_{n - 1} (\mu) + \mu \left (x_n - \frac{\mu}{2} \right) \right\}.
\end{equation}
We would then use $\max_{\mu} Q_n(\mu)$ as our test statistic.
It is straightforward to see that for any $\mu_1$, $Q_n(\mu_1)=\mathcal{Q}_{n,\mu_1}$. Thus if we can efficiently calculate { the function} $Q_n(\mu)$ then our test statistic is equivalent to the maximum value of the sequential-Page statistic over all possible choices of post-change mean. 
Furthermore, the following { proposition \cite[see, for instance, Example 2.4.3 in ][]{basseville1993detection} } shows that this test statistic is equivalent to the Page-CUSUM statistic (\ref{eq:page-CUSUM}), or equivalently the maximum of the MOSUM statistic (\ref{eq:MOSUM}) over all possible windows.
\begin{proposition} \label{prop1}
The maximum of $Q_n(\mu)$ satisfies
\[
\max_\mu Q_n(\mu)=\frac{1}{2}P(n)^2 = \frac{1}{2}\max_{w} M_w(n)^2,
\]
where $P(n)$ is the Page-CUSUM statistic and $M_w(n)$ is the MOSUM statistic with window size $w$.
\end{proposition}
The proof of this can be found in Appendix \ref{app:proof-prop1}. 

A description of the resulting algorithm for online changepoint detection is given in Algorithm \ref{alg:FOCuS_base_recursion}. We call this the Functional Online CuSUM (\Focus{}) algorithm. 
To be able to distinguish this version, that assumes a known pre-change mean, from the version we introduce in the next section, we call Algorithm \ref{alg:FOCuS_base_recursion} \FocusZ{}. The \FocusZ{} algorithm is only useful if it is computationally efficient, and in particular if we can implement Steps 1 and 2 efficiently. These steps correspond to solving the recursion in (\ref{eq:functional-recursion}) to obtain { the function} $Q_n(\mu)$ from $Q_{n-1}(\mu)$ and then maximising { the function} $Q_n(\mu)$. Below, we describe each of these steps in turn, and present results on their average computational cost.

\begin{algorithm}[tb]
	\caption{\FocusZ{} (one iteration)}
	\label{alg:FOCuS_base_recursion}
	\KwData{$x_n$ the data at time $n$; $Q_{n - 1} (\mu)$ the cost function from the previous iteration.}
	\KwIn{$\lambda > 0$}
    
    $Q_n(\mu) \arr \max \left\{ 0, \ Q_{n - 1} (\mu) + \mu \left (x_n - \frac{\mu}{2} \right) \right\}$ \tcp*{Algorithm \ref{alg:melkmans} : amortized $O(1)$}
    
    $\mathcal{Q}_n \arr \max_\mu Q_n(\mu)$ \tcp*{Theorem \ref{th:numberofchanges} : average $O(\log(n))$}
    

    \If{$\mathcal{Q}_n \geq \lambda $}{
    \Return{$n$ as a stopping point};
    }
    
    \Return{$Q_n(\mu)$ for the next iteration.}
\end{algorithm}

\subsubsection{Step 1: Updating the intervals and quadratics}
For Step 1 of Algorithm \ref{alg:FOCuS_base_recursion} we propose to update the { the function} $Q_n(\mu)$ separately for $\mu>0$ and $\mu<0$. These can be updated in an identical manner, so we will only describe the update for $\mu>0$. We will use the fact that (\ref{eq:functional-recursion}) maps piecewise quadratics to piecewise quadratics, and hence $Q_n(\mu)$ will be piecewise quadratic \cite[see][for a similar idea]{Maidstone} and can be stored as a list of ordered intervals of $\mu$ together with the coefficients of the quadratic for $Q_n(\mu)$ on that interval. Let $S_t=\sum_{j=1}^t x_j$ be the sum of the first $t$ data points. At time $n$ the quadratic introduced at iteration $\tau$ will be of the form
\begin{equation} \label{eq:summary}
\mu\left( \sum_{t=\tau+1}^n x_t- (n-\tau)\frac{\mu}{2}\right) = \mu\left( (S_n-S_\tau)- (n-\tau)\frac{\mu}{2}\right).
\end{equation}
Thus, if at time $n$ we know $n$ and $S_n$, its coefficients can be calculated if we store $\tau$ and $S_\tau$. This information stored for the quadratic does not need to be updated at each iteration.

{ So at any time $t$ we are able to produce a compact summary of the function $Q_t(\mu)$ by storing a set of triples $(\tau_i,s_i,l_i)$ for each of the $k_t$ quadratics that define the piecewise quadratic $Q_t(\mu)$. The entries are $\tau_i$, the time at which the $i$th quadratic is introduced, $S_i$ the sum of observations up to $\tau_i$, and $l_i$ the left-hand point of the interval of $\mu$ for which the $i$th quadratic is optimal. The quadratics are ordered so that $0=l_1<\cdots<l_k$, thus for $i<k$ the $i$th quadratic is optimal for $\mu\in[l_i,l_{i+1})$, with the $k$th quadratic optimal for $\mu \in [l_k, \infty).$ Furthermore, if $l_i<l_j$ then $\tau_i<\tau_j$, as quadratics introduced more recently will be optimal for larger values of $\mu$ than quadratics introduced less recently. (This can be shown as all quadratics go through the origin, and the $\mu^2$ coefficient of the $i$th quadratic will be larger than that of the $j$th quadratic.)

Now, consider an iteration to calculate the function $Q_{n}(\mu)$ from the function $Q_{n-1}(\mu)$. This is given in Algorithm \ref{alg:melkmans}. We will have currently stored the $k_{n-1}$ triples associated with the quadratics that define $Q_{n-1}(\mu)$. We split the recursion into two. First we calculate the intermediate function $Q^*_n(\mu)=Q_{n-1}(\mu)+\mu(x_n-\mu/2)$. This will update the coefficients of each of the $k_{n-1}$ quadratics that define $Q_{n-1}(\mu)$. However as we are defining the quadratics in terms of the summary statistics of the sum of the data points (\ref{eq:summary}), this is achieved by updating the sum of all data points $S_n=S_{n-1}+x_n$. 

Second we calculate $\max\{0,Q^*_n(\mu)\}$. To do this we first add a quadratic corresponding to the zero-line. This will have triple $(n,S_n,l)$, for some $l$ such that $Q^*_n(l)=0$, that we need to calculate. 

A key observation is that the difference between a quadratic introduced at iteration $\tau$ and the zero-line gives  that the zero-line is better 
on the interval 
\begin{equation}\label{eq:intervalFOCUS0}
\left[ 2\sum_{t=\tau+1}^n \frac{x_t}{(n-\tau)} \ , \ +\infty \right)
=
\left[ 2 \frac{S_n-S_\tau}{(n-\tau)} \ , \ +\infty \right).
\end{equation}
Considering all $\tau$ we get that the zero-line is better than all others on 
\begin{displaymath}
\left[ 2 \max_{\tau} \frac{S_n-S_\tau}{(n-\tau)}  \ , \ +\infty \right).
\end{displaymath}
Therefore we have that the final component for the triple defining the zero-line is
\begin{equation}\label{eq:simpleinterval2}
    l=2\max_{\tau} \frac{S_n-S_\tau}{(n-\tau)} . 
\end{equation}
If we can calculate $l$ then it immediately follows that any quadratic with $l_k>l$ can be removed. And any quadratic with $l_k<l$ will be unaffected (the interval on which they are optimal may be changed, but the triple that needs to be stored for the quadratic will be unaffected). By using the ordering of the quadratics mentioned above,  we find $l$ by comparing the zero-line with each of each quadratic in turn, starting with $k_{n-1}$th quadratic and stepping through them in decreasing order. If we are considering the $i$th quadratic we check whether $Q^*_{n}(l_i)<0$ or not. This is equivalent to checking whether the $i$th quadratic is less than 0 at $l_i$, or not. If it is we remove the quadratic and move to the $(i-1)$th quadratic (or stop if $i=1$). If $Q^*_{n}(l_i)>0$ then $l>l_i$ and we find $l$ as the positive value of $\mu$ such that the $i$th quadratic is equal to 0. 
}

\begin{algorithm}[!htb]
	\caption{Algorithm for $\max\{0, \ Q_{n-1}(\mu)+ \mu(x_n-\mu/2)\}$ for $\mu>0$}
	\label{alg:melkmans}
\KwData{$Q^+_{n} (\mu) = \mathrm{Q}$ an ordered set of triples $\{q_i = (\tau_i,\ s_i,\ l_i)\ \forall \ i = 1, \dots,\ k\}$, $x_{n} \mbox{ and } S_{n-1}$}
	
	$S_n \arr S_{n-1}+x_n$ \tcp*{update cumulative sum}
    $q_{k+1} \arr ( \tau_{k+1}=n,\ s_{k+1}=S_n,\ l_{k+1}=\infty)$  \tcp*{new quadratic}
    $i \arr k$;\\
	\While {$ 2 (s_{k+1} - s_i) - (\tau_{k+1} - \tau_i) l_i \leq 0$ \mbox{ and} $i\geq 1$}{
	    $i \arr i - 1$;\\
	}
    
    $l_{k+1} \arr  \max\{0, \ 2 (s_{k+1} - s_i) / (\tau_{k+1} - \tau_i)\} $\tcp*{update new border}
    
	\If{$i \neq k$} {
	    $\mathrm{Q} \arr \mathrm{Q} \setminus \{q_{i+1}, \ \dots, q_{k}\}$\tcp*{pruning old quadratic}
	} 

    \Return{$\{\mathrm{Q}, q_{k+1}\}$, ~ $S_n$}
\end{algorithm}

\begin{figure}
    \centering
    \includegraphics[scale=0.9]{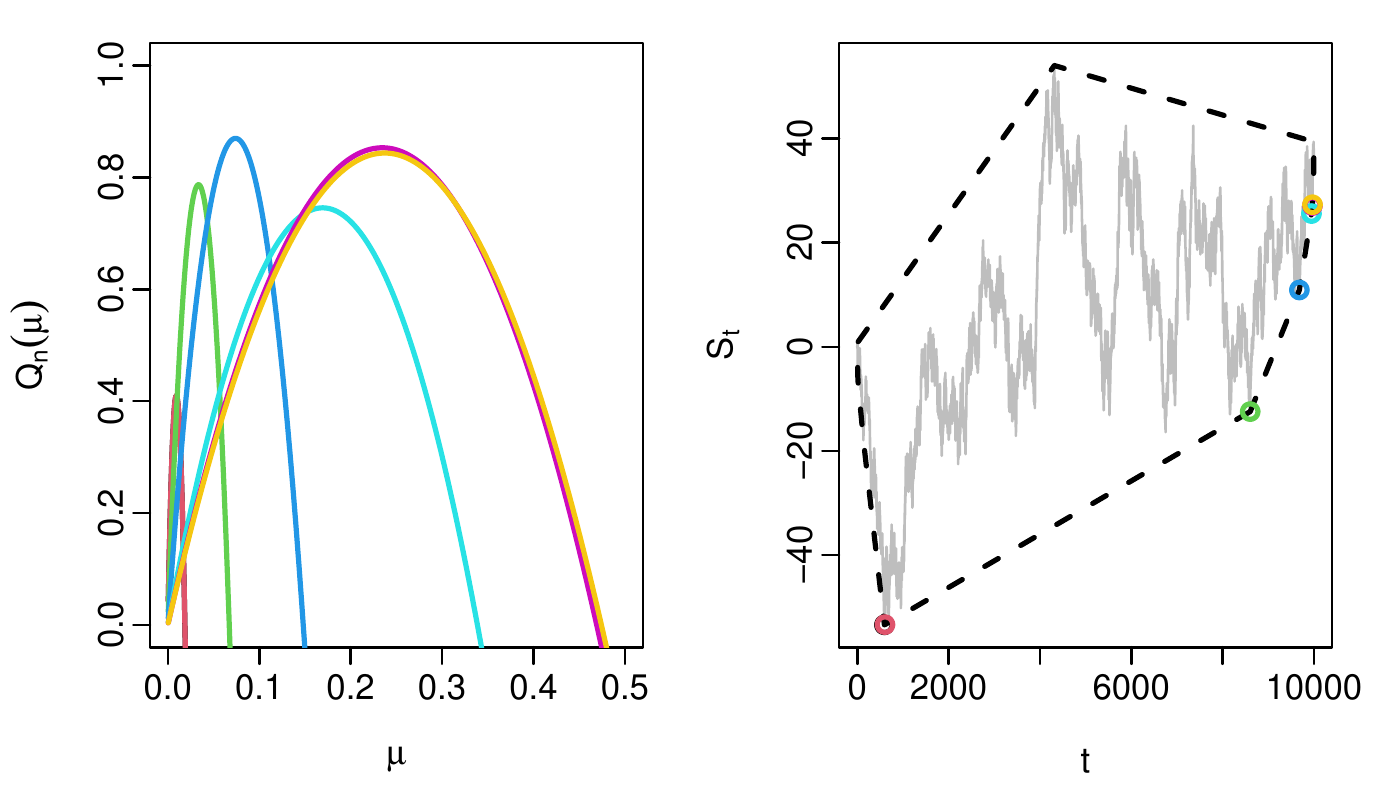}
    \caption{Example showing that the quadratics kept by \FocusZ{} are introduced at times related to vertices of the convex hull of the random walk $S_t=\sum_{i=1}^t x_i$. (Left) Plot of quadratics defining $Q_{n}(\mu)$. (Right) Plot of $S_t$ as a function of $t$ (grey), and the convex hull of the points $(t,S_t)$ (dashed line). We have circled all points $(t,S_t)$ associated with the time each quadratic in the left-hand plot was introduced (with the same colouring), and these correspond to vertices of the convex hull. All vertices lie on the lower convex minorant because we are considering a positive change in mean. We only keep quadratics associated with vertices to the left of sides of the convex hull with positive gradient due to the condition that the post-change mean is larger than the pre-change mean of 0. Changing the pre-change mean thus affects only which of the vertices on the lower minorant correspond to quadratics that are kept by \FocusZ{}}
    \label{fig:chull}
\end{figure}

{ 

\begin{figure}
    \centering
    \includegraphics[scale=0.9]{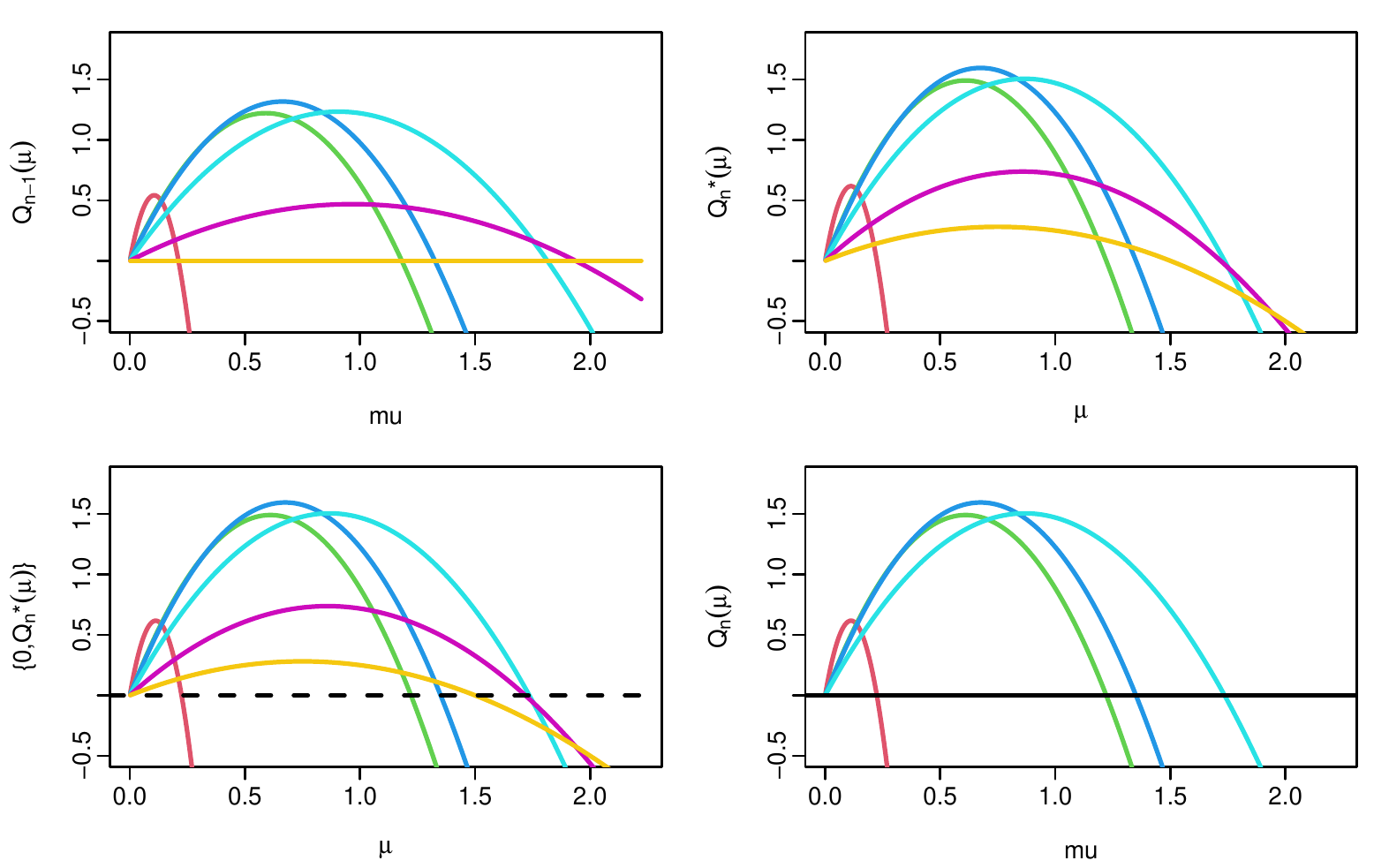}
    \caption{Example of one iteration of \FocusZ{}. (Top left) The output at time $n-1$ is the function $Q_{n-1}(\mu)$ which is the maximum of a set of quadratics. In this example it is the maximum of 6 quadratics, one of which is the zero-line. In the algorithm each quadratic is represented by a triple $(\tau,s,l)$, which are the time the quadratics is introduced, the sum of observations at that time, and the smallest $\mu$ value for which it is optimal. A key property is that these are ordered so the quadratics introduced early have a higher coefficient of $\mu^2$ and are thus optimal for lower values of $\mu$. (Top right) The function $Q^*_n(\mu)=Q_{n-1}(\mu)+\mu(x_n-\mu/2)$. This is still the maximum of six quadratics, but the co-efficients have changed from the top left plot. In practice this involves no need to update the triples for the quadratics. (Bottom left) We introduce the zero-line (black dashed line), which is the quadratic introduced at time $n$. We compare the zero-line with each current quadratic by seeing which is larger at the lowest $\mu$ value for which the quadratic was optimal, starting with the most recent quadratics (i.e. yellow, then purple etc.). If the quadratic is below the zero-line it is no longer optimal and can be removed. For the first quadratic that is above the zero-line, we calculate the non-zero value of $\mu$ where the quadratic and zero-line intercept. The maximum of this and 0 is the smallest $\mu$ value for which the zero-line is optimal. (Bottom right) The function $Q_n(\mu)$ after removing the two quadratics that are no longer optimal. }
    \label{fig:FOCUSexample}
\end{figure}

A pictorial representation of the algorithm is shown in Figure \ref{fig:FOCUSexample}. The condition for $l$ is related to the slope of the random walk $S_t$, $t=1,\ldots,n$. As shown for an example in Figure \ref{fig:chull}, the quadratics that we keep are related to a subset of the vertices of the lower convex hull of this random walk (this property is proven formally in the proof of Theorem \ref{th:numberofchanges}), a property that will be useful in bounding the computational complexity of the algorithm. Furthermore Algorithm \ref{alg:melkmans} is closely related to Melkman's algorithm for finding the convex hull of a set of points \citep{melkman1987line}.
}

We can show that Algorithm \ref{alg:melkmans} has an amortized 
per-iteration cost that is $O(1)$. The intuition is that each quadratic is added once and removed once, and otherwise unchanged. Thus the average per-iteration cost is essentially the cost of adding and of removing a quadratic.

\begin{theorem} \label{thm:complexity_step1}
The worst case complexity of Algorithm \ref{alg:melkmans} 
for any data $x_1,\ldots, x_T$ is $O(T)$ and its amortized complexity per iteration is $O(1)$. 
\end{theorem}

{\bf Proof:}
     At iteration $n$, let $k_n$ be the number of quadratics input, and let $c_n$ be the number of times the { while statement is evaluated.}   Let $C_1$ be the cost of steps 1 to 3 and 7 to 11, and $C_2$ be the cost of one set of one evaluation of steps 4 to 6. Then the computational cost of one iteration of Algorithm~\ref{alg:melkmans} is $C_1+c_n\times C_2$.
    
    The key observation is that 
    $k_{n+1}=k_{n}-(c_n-1)+1.$ 
    That is if we repeat Steps 4 to 6 $c_n$ times then we will remove $c_n-1$ quadratic in Step 9 and add one quadratic in Step~11. Furthermore $k_1=0$ and $k_{T+1}$ is the number of quadratics for $Q_T(\mu)$.
    Thus the total computational cost is
    \[
    \sum_{n=1}^T (C_1+c_n C_2) = C_1 T + C_2 \sum_{n=1}^T c_n = C_1 T + C_2 
    \sum_{n=1}^T (2+k_n-k_{n+1}) .
    \]
    Due to the cancellations in  the telescoping sum and the fact that $k_1=0$ we have 
    \[
    \sum_{n=1}^T c_n = 2T-k_{T+1} \leq 2T
    \]
    Thus the theorem holds 
    \hfill $\Box$

\paragraph{} As the overall computational cost is linear in $T$, the expected cost per iteration must be constant. The proof gives a form for the overhead in terms of the operations in Algorithm \ref{alg:melkmans}. In practice this cost is observed to be negligible relative to the cost of step 2 of Algorithm~\ref{alg:FOCuS_base_recursion}, namely that of maximising $Q_n(\mu)$.

\subsubsection{Step 2 : Maximisation}
To implement Step 2 of Algorithm \ref{alg:FOCuS_base_recursion} we first use the trivial observation that if $x_n>0$ then $Q_n(\mu)<Q_{n-1}(\mu)$ for all $\mu<0$. Thus to check if $\max_{\mu} Q_n(\mu) \geq \lambda$ we need only check this for $\mu>0$. Similarly if $x_n<0$ then we need only check for $\mu<0$. To perform the check we just loop over all quadratics stored for either $\mu>0$ or $\mu<0$, and for each one check if its maximum is greater than $\lambda$. For a quadratic with stored triplet $(\tau,s,l)$ this involves checking whether 
\begin{equation} \label{eq:quadmax}
(S_n-s)^2 \geq 2\lambda(n-\tau).
\end{equation}
If we flag a change at time $n$, then we can also output the value of $\tau$ corresponding to the quadratic whose maximum is largest, and this will be an estimate of the time of the change.
The computational cost is thus proportional to number of quadratics that are stored, and can be bounded using the following result.

\begin{theorem}
\label{eq:FOCuS0_n_intervals}
Let $x_1, \dots, x_T, \dots$ be a realization of the process $X_i = \mu_i + \epsilon_i$ where $\epsilon_i $ are independent, identically distributed continuous random variables with mean 0. Let the number of quadratics stored by \FocusZ{} for $\mu>0$ at iteration $T$ be 
$\#\mathcal{I}^0_{1:T}$. Then if $\mu_i$ is constant
\[
E(\#\mathcal{I}^0_{1:T}) \leq (\log(T)+1),
\]
while if $\mu_i$ has one change prior to $T$ then
\[
E(\#\mathcal{I}^0_{1:T}) \leq 2(\log(T/2)+1).
\]
\end{theorem}
The proof of this can be found in Appendix \ref{app:number_of_ints}. { The key idea is to use one-to-one correspondence between each quadratic we need to keep and the vertices of the convex hull of the random walk $S_t=\sum_{i=1}^t x_i$ that is shown in Figure \ref{fig:chull}. Standard results \citep{andersen1955fluctuations} give bounds on the vertices of a convex hull of a random walk where the increments are exchangeable.}

By symmetry, the same result holds for the number of quadratics stored for $\mu<0$. The conditions of the data generating mechanism are weak -- as the distribution of the noise can be any continuous distribution providing the noise is independent. The theorem shows that the expected per-iteration  time and memory complexity of \FocusZ{} at time $T$ is $O(\log T)$. Furthermore, the expected per-iteration cost is essentially equal to checking (\ref{eq:quadmax}) $\log(n)+1$ times if there has not been a change, and for $2(\log(n/2)+1)$ if there has been an, as yet, undetected change. A change of fixed size is detected in $O(1)$ iterations, and thus the overall computational time, for large $T$, will be dominated by the cost of iterations prior to the changepoint. For data of size one million, the bound on the number of quadratics is less than 15. { For \FocusZ{}, this bound can be improved upon. The bound is based on all quadratics that are stored corresponding to vertices on the convex minorant of the than random walk $S_t=\sum_{i=1}^t x_i$, but as shown in Figure \ref{fig:chull}, not all vertices on the convex minorant correspond to quadratics that are stored. In fact, a simple symmetry argument suggests that only half of them do. This is validated empirically in Appendix \ref{sec:FOCuSTheoEMPIRICAL}.}

The \FocusZ{} algorithm is not strictly online, due to the cost per iteration not being bounded. But it is simple to introduce a minor approximation that is online. Assume we have a constraint that means we can find the maximum of at most $P$ quadratics per iteration. A simple approximation is to introduce a grid of points $\pm m_p$ for $m_p \in \mathbb{R}^+, \ p = 1, ... P$. There are then two natural approaches. One is that if we have $P+1$ quadratics stored we prune to $P$ quadratics by removing the first quadratic whose interval does not contain a grid point. Alternatively we can keep all quadratics but only find the maximum of the quadratics whose interval contains a grid point. The advantage of this latter approach is that it avoids any approximations to { the function} $Q_n(\mu)$ which could propagate to future values of { the functions} $Q_t(\mu)$ for $t>n$. Both these methods would dominate using the sequential-Page approach that used the same grid for $\mu_1$ values. For example, if $\tilde{Q}_n(\mu)$ denotes the approximation to $Q_n(\mu)$ using the first approach, then we have $\tilde{Q}_n(\mu_1)=\mathcal{Q}_{n,\mu_1}$ for all $\mu_1$ in our grid.
{ Thus, for either method, the maximum value of the quadratics evaluated by \FocusZ{} will be equal to or greater than the maximum of the sequential-Page statistics at the grid points. As with sequential-Page, it is sensible to use a geometric scaling for the grid points $\pm m_p$ (see below). }


\input{sim_study_v3}


\section{Unknown pre-change mean}\label{sec:pre-change-unknown}

{
Assume we are observing a sequence of observations $x_1, \dots, x_{n}$ distributed as a $N(\mu_0, \sigma)$ prior to any change, and as $N(\mu_1, \sigma)$ after the change, with $\sigma$ known and $\mu_1 \neq \mu_0$. As before, without loss of generality we will assume $\sigma=1$. As suggested by \cite{yu2020note}, we can base a test for a change on the likelihood ratio statistic. 
\begin{equation}
LR_n=
 \underset{\substack{\tau \in \{ 1, \dots, n-1 \} \\ \mu_0, \mu_1 \in \mathbb{R}} }{\max} \left\{ - \sum_{t=1}^{\tau} (x_t - \mu_0)^2 - \sum_{t=\tau+1}^n (x_t - \mu_1)^2 \right\}
 -  \max_{\mu \in \mathbb{R}} \left\{ - \sum_{t=1}^n (x_t - \mu)^2\right\}.
\end{equation}
\cite{yu2020note} present finite-sample results that demonstrate the statistical optimality of such a test.  They also present algorithms for evaluating this test statistic. Their fastest algorithm that avoids any approximation is  $O(n)$ in computational complexity per iteration while being  $O(n)$ in storage, which make their methodology infeasible to a true online setting. For the rest of this paper will refer to that algorithm as Yu-CUSUM.

To be consistent with, and show the links to, how we calculated the test statistic in the pre-change mean known case, we introduce the following functions, 
\[
\mathcal{Q}_{\tau,n}(\mu_0,\mu_1) = 
- \frac{1}{2}\sum_{t=1}^\tau (x_t-\mu_0)^2 - \frac{1}{2}\sum_{t=\tau+ 1}^n (x_t-\mu_1)^2.
\]
This is the log-likelihood for a model with a change at $\tau$ from mean $\mu_0$ to $\mu_1$. If $\mu_1=\mu_0$ this reduces to the log-likelihood for the data having mean $\mu_0$ throughout.

Let $\mathcal{Q}_{n}(\mu_0,\mu_1)=\max_{\tau\in {1,\ldots,n-1}} \mathcal{Q}_{\tau,n}(\mu_0,\mu_1)$.
Then the log likelihood-ratio statistic can be calculated as
\[
LR_n =  2\left\{\underset{\substack{ \mu_0, \mu_1 \in \mathbb{R}} }{\max} \mathcal{Q}_{n}(\mu_0,\mu_1) - \max_{\mu_0 \in \mathbb{R}} \mathcal{Q}_{n}(\mu_0,\mu_0)\right\}.
\]
Thus if we can calculate the function $\mathcal{Q}_{n}(\mu_0,\mu_1)$ we can calculate the log likelihood-ratio test statistic.

If we fix $\mu_0$ and consider $\mathcal{Q}_{n}(\mu_0,\mu_1)$ as a function of $\mu_1$ only, then, up to a constant, this is the function calculated in the previous section. We can calculate this separately for $\mu_1>\mu_0$ and $\mu_1<\mu_0$, and in each case that function will be piecewise quadratic, with the quadratics introduced at times that are on the convex hull of $S_t=\sum_{i=1}^t x_i$ (see Figure \ref{fig:chull}). For a quadratic introduced at time $\tau$, the quadratic is of the form
\[
-\frac{1}{2}\sum_{t=1}^n x_t^2 + \tau \mu_0\left(\frac{S_\tau}{\tau}-\frac{\mu_0}{2} \right) + (n-\tau)\mu_1
\left(\frac{S_n-S_\tau}{n-\tau}-\frac{\mu_1}{2} \right).
\] 
Thus if we consider $\mu_1>\mu_0$, say, then the the function $\mathcal{Q}_{n}(\mu_0,\mu_1)$ will also be piecewise quadratic, with one quadratic corresponding to each point on the convex minorant of $S_t$. This is shown pictorially in Figure \ref{fig:chull2}.

\begin{figure}
    \centering
    \includegraphics[scale=0.9]{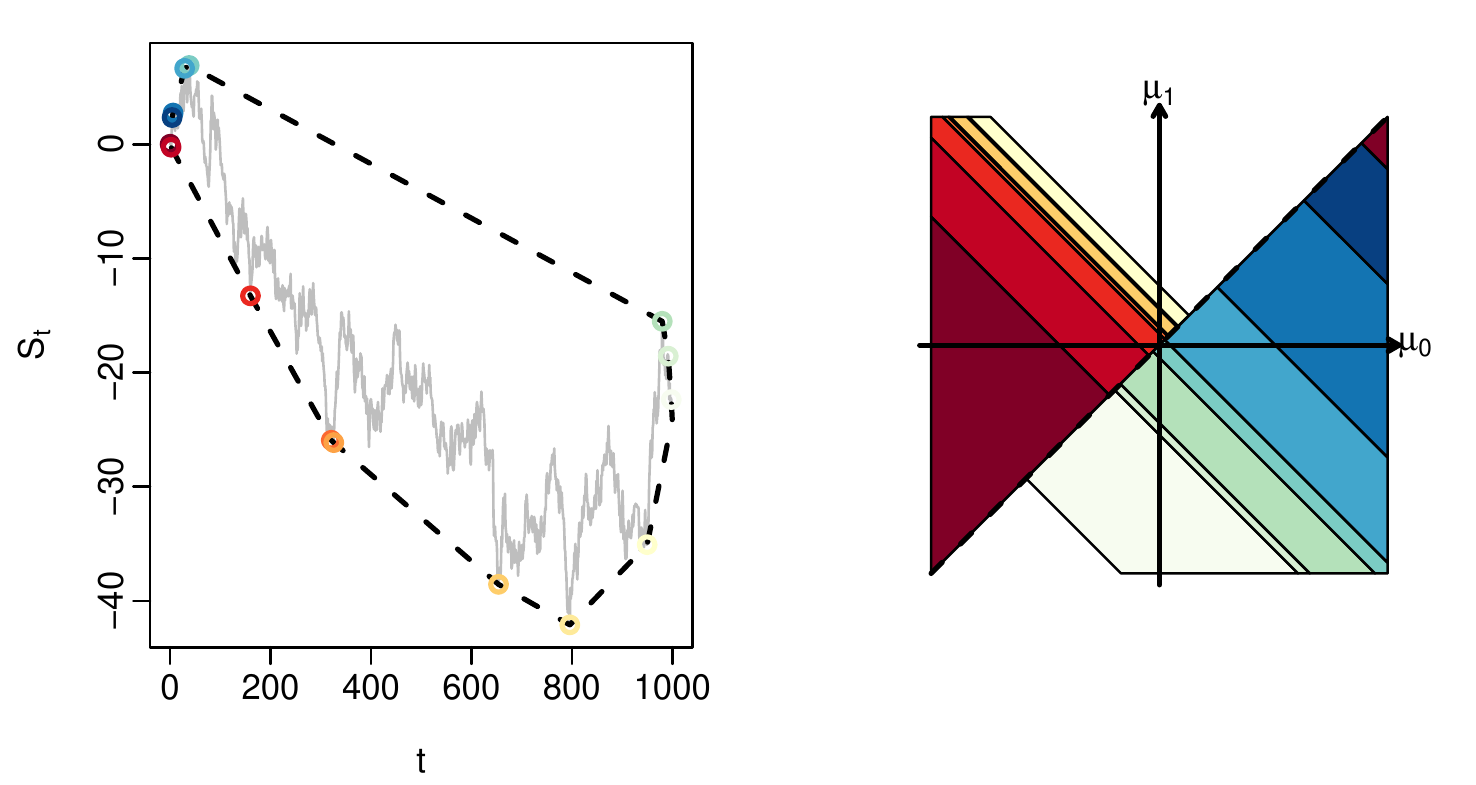}
    \caption{Convex hull of the points $(t,S_t)$ where $S_t=\sum_{i=1}^t x_i$ (left-hand plot), and plot of regions in $(\mu_0,\mu_1)$ space for which a different quadratic is optimal for $\mathcal{Q}_n(\mu_0,\mu_1)$ (right-hand plot). Each quadratic in the definition of $\mathcal{Q}_n(\mu_0,\mu_1)$ corresponds to a vertex on the convex hull of the random walk of the data (in that is corresponds to a change at the time associated with that vertex) and is shaded in the same colour as the corresponding vertex is labelled. }
    \label{fig:chull2}
\end{figure}

An algorithm to calculate the log likelihood-ratio statistic is thus similar to that of the previous section, except for two small differences.
\begin{enumerate}
    \item For the interval update (step 1), in \FocusZ{} for up-changes we could restrict our attention to $\mu_1 \in [\mu_0, +\infty)$ (resp. $(-\infty, \mu_0]$ for down-changes), whereas in \Focus{}, as we do not know the value of the first segment mean, we need to consider all cases for $m$, i.e. $m \in (-\infty, +\infty).$ This means that for \Focus{} and up-changes we run Algorithm 2 with line 7 changed to
    \[
    l_{k+1} \arr  2 (s_{k+1} - s_i) / (\tau_{k+1} - \tau_i).
    \]
    That is we no longer take the maximum of this and the pre-change mean value. For down-changes we can apply the same algorithm but, by symmetry for data with the sign flipped, i.e. $-x_{1:n}$. 
    \item For the maximisation (step 2) in \FocusZ{} we only need to optimize the value of the last segment, whereas in \Focus{} we also need to optimize over the pre-change mean. That is, at time $n$, for a quadratic defined by triple $(\tau_i,s_i,l_i)$ we calculate
    \[
    n\left(\frac{S_n}{n}\right)^2 - \tau_i\left(\frac{S_i}{\tau_i}\right)^2 - (n-\tau_i)\left(\frac{S_n-S_i}{n-\tau_i}\right)^2.
    \]
    We then find the maximum value of these values, maximising over all quadratics stored at time $n$.
\end{enumerate}
}

By the same argument as that of Theorem \ref{thm:complexity_step1}, we have that solving of the recursion has an average cost per iteration that is constant. We derive in Theorem \ref{th:numberofchanges} in Appendix \ref{app:number_of_ints} the same bound on the expected number of candidates as for \FocusZ{}, showing that  the  expected  per-iteration  time  and  memory  complexity  of maximising the solution of \Focus{} at time $T$ is $O(\log(T))$. 

\subsection{Simulation Study}

\begin{figure}
    \centering
    \begin{subfigure}{0.32\linewidth}
        \centering
        \includegraphics[width=\linewidth]{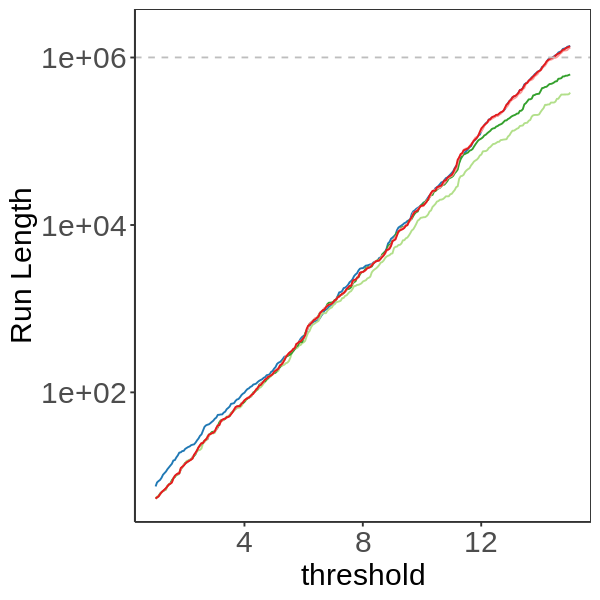}
    \end{subfigure}
    \begin{subfigure}{0.32\linewidth}
        \centering
        \includegraphics[width=\linewidth]{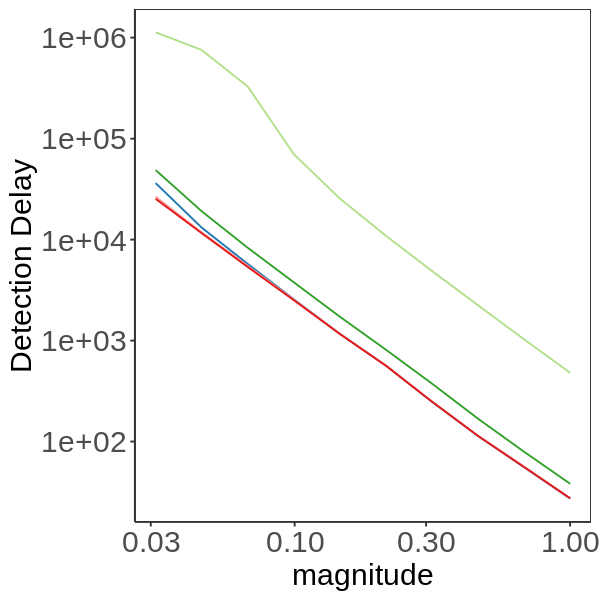}
    \end{subfigure}
    \begin{subfigure}{0.32\linewidth}
        \centering
        \includegraphics[width=\linewidth]{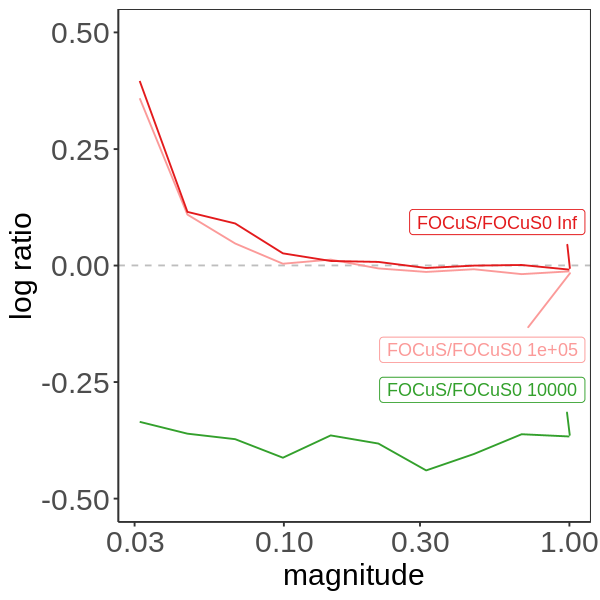}
    \end{subfigure}
    \caption{Comparison between \Focus{} pre-change unknown and pre-change known: Average run length against threshold (left); average detection delay against magnitude of change (middle); and log-ratio of average detection delay of pairs of methods against magnitude of change (right). For the first two plots: the methods are \Focus{} (blue); \FocusZ{} with pre-change mean known (red); and  \FocusZ{} with different training data sizes: 1000  (light green),  $1\e{4}$ (dark green) and $1\e{5}$ (pink).}
    \label{fig:sim-unknown-mean}
\end{figure}

We make a comparison between \Focus{} with the pre-change mean unknown and \FocusZ{} with the pre-change-mean known learned over a training sequence. This is because, as mentioned in the introduction, one could estimate the mean of a Gaussian process when the pre-change mean is unknown, and use such a value to run the algorithms introduced in Section \ref{sec:pre-change-known}. We study in particular the performances of \FocusZ{} as we vary the size of training data from $1000$ observations up to $1\e{5}$. As a benchmark we also compare to \FocusZ{} with known pre-change mean -- which should show the best possible performance.
We compare both average run-length as a function of the threshold, and detection delay as a function of the magnitude of a change. For each experiment, we report summaries over 100 replicates, and the results on detection delay are for thresholds chosen so each algorithm has an average run-length of $1\e{6}$. In all cases we simulate data with $1\e{5}$ data points prior to the change. 
Results are summarised in Figure \ref{fig:sim-unknown-mean}.

We note how \Focus{} requires a smaller threshold to achieve the same average run length of various \FocusZ{} implementations, however differences become negligible when comparing the methods over larger training sizes. Concerning detection delay, the advantage of \Focus{} is that it can improve its estimate of the pre-change mean using the data prior to any change, thus we see substantial benefits of \Focus{} relative to \FocusZ{} when the amount of training data is small. However, when the amount of training data is of the same order as the amount of data prior to the change, \FocusZ{} has more power than \Focus{} for detecting very small changes, whereas \Focus{} has more slightly more power for detecting larger changes.
The fact that \Focus{} is worse than \FocusZ{} with a large training dataset for small magnitudes is intuitively expected. On the one hand, the cost optimized by \Focus{} is bounded making small changes undetectable (the cost will always be smaller than the pre-defined threshold); this is the cause for the detection threshold of \cite{yu2020note}. On the other hand, the cost of \FocusZ{} is unbounded, therefore assuming we have a reasonably close estimate of the pre-change mean small changes can be identified.

\section{Extensions of \Focus{}}\label{sec:focus_extensions}

\subsection{\Focus{} in the presence of outliers}

Further extensions of \Focus{} are to use different loss functions to the square error loss obtained from a Gaussian log-likelihood. Motivated by the application in Section \ref{sec:application} we will consider a robust loss function, the biweight loss, which enables us to detect changepoints in the presence of outliers \cite[see][]{Fearnhead} { (though other loss functions that are piecewise quadratic, such as $L_1$ loss can be used with a similar approach)}. This loss is just the square error loss but capped at a maximum value, $K$, chosen by the user. To be consistent with earlier section, we can then define an associated measure of a fit to the data, as minus this loss,
\begin{equation}\label{eq:LR_robust}
    F(x_t, \mu_1) = -\min\left\{ \left(\frac{\mu_1}{2} - x_t \right)^2, K \right\}.
\end{equation}
We then aim to detect a change by monitoring the resulting fit to the data if we maximise over potential locations of a change. This leads to the following functional recursion:
\begin{equation}
        \label{eq:functional-recursion_robust}
        Q_n(\mu) = \max \left\{ { \max_{\mu_0}} \sum_{t=1}^{n} F(x_t, \mu_0), \ Q_{n - 1} (\mu) + F(x_n, \mu) \right\}.
\end{equation}
Using ideas described in Section 3.2 of \cite{Fearnhead} it is straightforward to implement this recursion for all $\mu$ efficiently. For this model we are unable to recover a bound on the expected number of candidate changepoints. However we observed empirically that the cost for iteration $n$ is in $O(\log(n))$ (see Figure \ref{fig:timing-pre-change-ukn}).

\subsection{Simulation Study}\label{sec:sim-study-unknown}
 
\begin{figure}[tb]
    \centering
    \includegraphics[width=.55\linewidth]{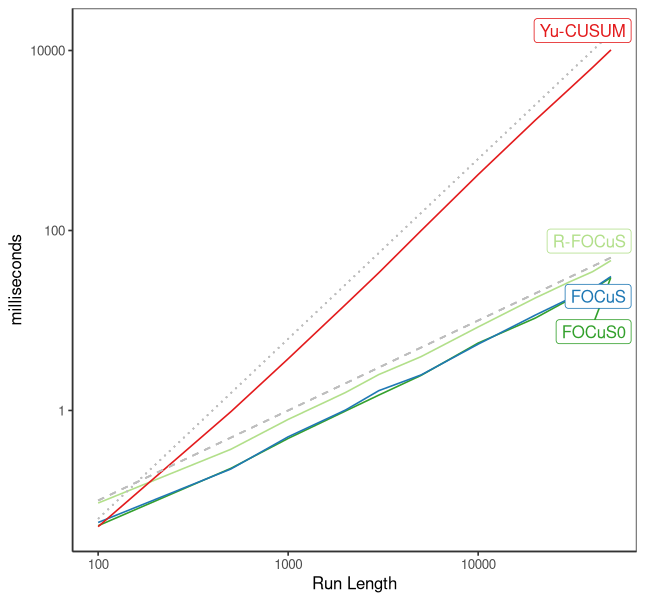}
    \caption{Runtime in milliseconds of \FocusZ{}, \Focus{}, R-FOCuS and Yu-CUSUM in function of the length of the sequence (log-scale on both axes). Grey lines refer to an expected $O(n)$ increase (dashed) and $O(n^2)$ increase (dotted).}
    \label{fig:timing-pre-change-ukn}
\end{figure}

In Figure \ref{fig:timing-pre-change-ukn} we give a comparison of the runtime between \FocusZ{}, \Focus{}, the robust implementation introduced in \eqref{eq:functional-recursion_robust} (R-FOCuS) and Algorithm 3 from \cite{yu2020note}, denoted as Yu-CUSUM. Runtimes were recorded for multiple finite sequences of lengths ranging from $100$ to $5\e{4}$. To produce a fair comparison both implementations were written in C++, all simulations were performed on a common personal computer. We find little difference when comparing \FocusZ{} with \Focus{}, both showing an empirical linear increase in timings with the latter being slightly slower. When comparing \Focus{} to Yu-CUSUM, we find a comparable runtime only up to $n = 100$, after which \Focus{} is faster, due to the quadratic computational complexity of Yu-CUSUM. Lastly, we notice how R-FOCUS, while still retaining a linear computational complexity, has a larger overhead compared to the simpler implementations.

\section{Application of \Focus{} to the AWS Cloudwatch CPU utilization} \label{sec:application}

\begin{sidewaysfigure}
    \centering
    
    \includegraphics[width=\linewidth]{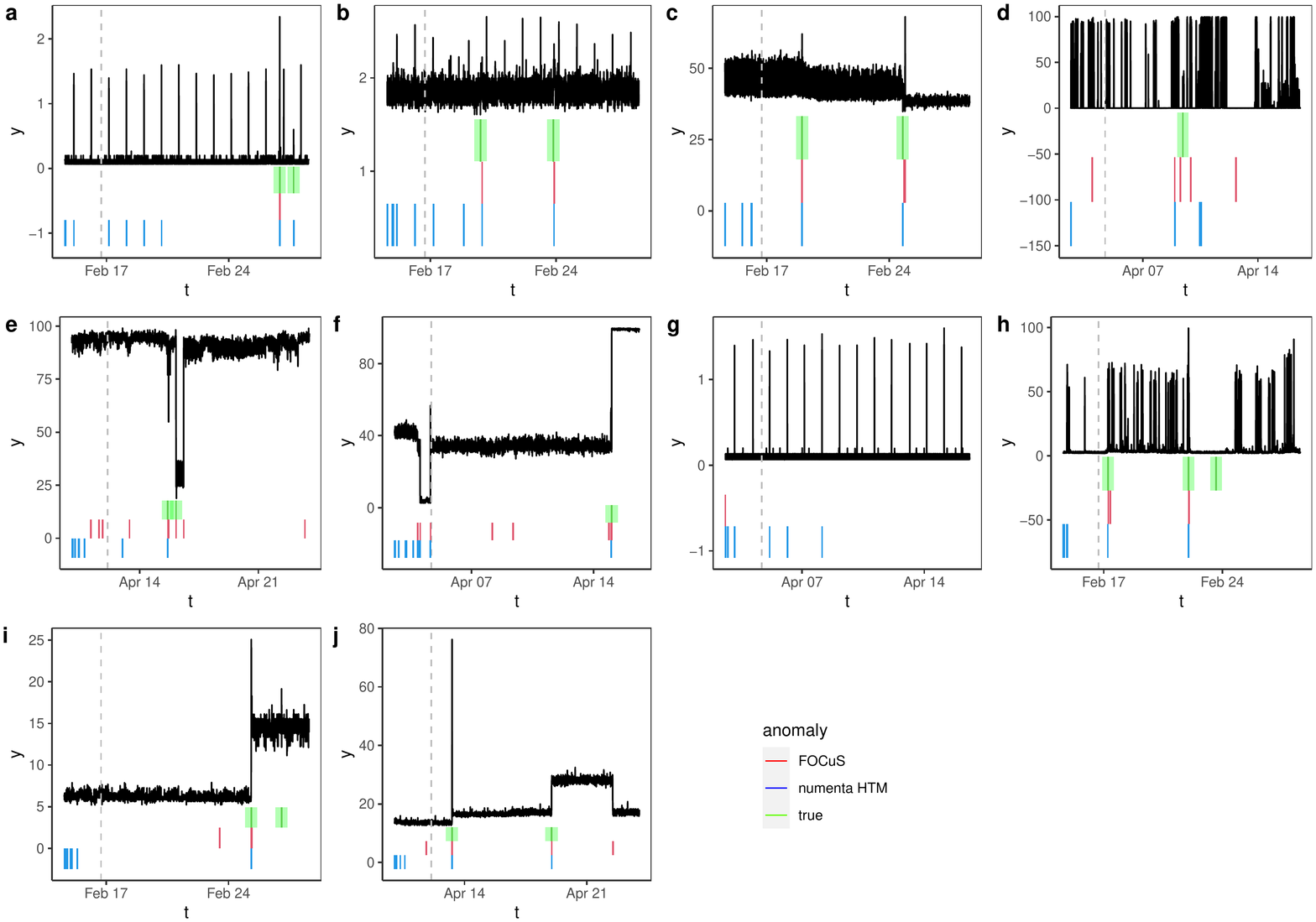}
    {\caption{The 8 (\textbf{a} - \textbf{j}) different time series of AWS Cloudwatch CPU utilization. Green, red and blue segments correspond, respectively, to the real and estimated anomaly locations of \RFocus{} and Numenta HTM. The green rectangle around each anomaly is the anomaly window (the area in which an anomaly must be detected to count as a true positive). Multiple detections within the green lines are allowed, and are not considered as false positives. The dashed line corresponds to the probation period, used for training of tuning parameters: detections before the dashed line are not accounted for in the final result.} \label{fig:real_data}}
\end{sidewaysfigure}

We now evaluate \Focus{} by comparing with a bespoke anomaly detection algorithm on the Amazon CPU utilization datasets from the the Numenta Anomaly Benchmark \cite[]{ahmad2017unsupervised}. The aim with these datasets is to detect anomalous behaviours in the CPU utilization of various Amazon Cloudwatch instances. For each dataset anomalous behaviours have been manually flagged by experts, and those stand as the ground truth. The data sets are shown in Figure \ref{fig:real_data}, and demonstrate a range of behaviour. As point anomalies are common we will use the \RFocus{} algorithm.

When evaluating algorithms we will follow the methodology in \cite{ahmad2017unsupervised}. A detection is deemed to be correct if it lies within $\pm 0.05 \cdot n$ of the true anomaly, where $n$ is the length of the time series; and multiple detection within the window are allowed. A method can use the first  $15\%$ of each dataset, a portion of data known to not include any anomalies, to set tuning parameters. We use this data to tune both $K$ in the biweight loss and the detection threshold as described in Appendix \ref{sec:param-est}.

As some data sets have multiple anomalies to be detected, we have to adapt \RFocus{} so that it does not stop once a change to some anomalous behaviour has occurred. To adapt \RFocus{} we simply initiate the procedure again at the estimated changepoint location after a detection is triggered. In order to reduce the number of false positives and to extend the average run length of the algorithm, at each detection we inflate the threshold by a factor of $\log(\tau_s) / \log(\tau_s - \tau_{s-1})$, with $\tau_0, ..., \tau_k$ being a vector of estimated changepoint locations. { This is the same inflation as used during the training phased, and its form is based on theory that suggest the threshold should be proportional to the log average run length \citep{yu2020note}. Inflating the penalty during the test data is conservative, as the changes may be real rather than false positives, but avoids issues with bursts of false positives caused by heterogeneity in the data.}

 We compare \RFocus{} with numenta HTM, the best performing algorithm to date on these data. Numenta HTM \cite[]{ahmad2017unsupervised} is an anomaly detection algorithm that employs an unsupervised neural network model to work with temporal data \citep{cui2016continuous} to perform anomaly detection. 

Results are summarised in Figure \ref{fig:real_data} and Table \ref{tab:nab_score}.
We find that \RFocus{} has better performances in term of Precision, the proportion of true anomalies detected, and Recall, the proportion of detections that are true anomalies, compared with Numenta HTM. 
On a case to case basis, in most of the sequences both algorithms flagged correctly the anomalies. HTM overall achieves slightly shorter detection delays (with the exception of \textbf{f}), however it produces more false positives (13 false detections against 7 of \RFocus{}). In terms of missed detections, both algorithms perform similarly, with \RFocus{} missing an anomaly flagged by HTM in \textbf{a}, and HTM missing an anomaly flagged by \RFocus{} in \textbf{d}.

Overall, this shows the flexibility of \RFocus{} for online change detection, especially considering that \RFocus{} is a simpler approach which is operating under model misspecification, and with significantly shorter computational run-times than Numenta HTM (on 3000 observations \RFocus{} takes roughly 2 milliseconds against the 4 minutes for HTM). 

\begin{table}[]
\centering
\begin{tabular}{@{}lcc@{}}
\toprule
Detector    & Precision & Recall  \\ \midrule
\RFocus       & 0.58           & 0.82          \\
Numenta HTM & 0.50           & 0.76      
\end{tabular}
\caption{Precision and  Recall  for \RFocus{} and Numenta HTM.}
\label{tab:nab_score}
\end{table}

\section{Discussion}
\label{sec:discussion}

{
\begin{table}[]
\resizebox{\textwidth}{!}{%
\begin{tabular}{cc|cc|cc}
         &           & \multicolumn{2}{c|}{Pre-change mean known} & \multicolumn{2}{c}{Pre-change mean unknown} \\ \hline
sparsity & magnitude & {\texttt{ocd}}              & \FocusZ{}          & {\texttt{ocd}}              & \Focus                \\ \hline
0.01     & 0.25      & \textbf{318.30}      & 365.03              & \textbf{1281.24}     & 1446.36              \\
0.01     & 0.5       & \textbf{95.01}       & 95.37               & 336.23               & \textbf{121.29}      \\
0.01     & 1         & 36.37                & \textbf{25.15}      & 131.13               & \textbf{26.63}       \\
0.01     & 2         & 12.65                & \textbf{7.34}       & 59.83                & \textbf{7.38}        \\
0.05     & 0.25      & \textbf{503.46}      & 702.48              & \textbf{2164.01}     & 2172.04              \\
0.05     & 0.5       & \textbf{144.57}      & 180.37              & 622.55               & \textbf{317.72}      \\
0.05     & 1         & 43.91                & \textbf{43.38}      & 206.59               & \textbf{48.66}       \\
0.05     & 2         & 14.08                & \textbf{12.40}      & 86.05                & \textbf{12.77}       \\
0.1      & 0.25      & \textbf{616.11}      & 803.43              & 2348.26              & \textbf{2312.23}     \\
0.1      & 0.5       & \textbf{171.11}      & 233.34              & 823.61               & \textbf{525.72}      \\
0.1      & 1         & \textbf{50.88}       & 59.57               & 258.74               & \textbf{69.33}       \\
0.1      & 2         & \textbf{14.73}       & 16.99               & 106.61               & \textbf{17.64}       \\
1        & 0.25      & \textbf{904.74}      & 1110.48             & 2903.23              & \textbf{2506.79}     \\
1        & 0.5       & \textbf{245.00}      & 420.56              & 1769.32              & \textbf{1081.09}     \\
1        & 1         & \textbf{64.39}       & 130.30              & 675.58               & \textbf{190.29}      \\
1        & 2         & \textbf{17.85}       & 39.74               & 210.92               & \textbf{44.09}      
\end{tabular}%
}
\caption{Average detection delay for changes of different sparsity levels and magnitudes in multivariate sequences. We consider two scenarios, where the pre-change mean is known and the pre-change mean is unknown. For the former we use \FocusZ{}, for the latter we use \Focus{} and implement \texttt{ocd} with pre-change means estimated from training data of length 500. The sparsity specifies the proportion of the 100 data streams that change. For a change which affects $k$ series we simulate $Z_1,\ldots,Z_k$ independent standard normal variables, and change the mean of the $i$th series, for $i=1,\ldots,k$, by $m Z_i/\sqrt{\sum_{j=1}^k Z_j^2}$, where $m$ is the magnitude of the change. All changes occur at time 200.}
\label{tab:multiv-sim}

\end{table}
}

As shown in Section \ref{sec:pre-change-unknown}, one of the advantages of the functional pruning recursion lies in the numerous extensions possible when manipulating the cost directly. For instance, conditions on the underlying mean object of inference could be implemented to produce inference constrained to specific change patterns \citep{hocking2020constrained, runge2020gfpop}, or to account for fluctuating signals and autocorrelation in the noise \citep{romano2020detecting}, or allow for known behaviour of the the mean between changes \citep{jewell2020fast}.

The main limitation of the recursions are that they rely on functional pruning ideas that currently only work for functions of a univariate parameter \cite[though see][for ideas on extending pruning to higher dimensions]{runge2020finite}. However there is still potential for applying versions of \Focus{} in situations where multiple parameters may change, for example by separately testing for changes in each parameter and merging this information. { A simple way of merging the \Focus{} statistics is to take either their maximum or their sum \cite[see e.g.][]{mei2010efficient}. The maximum is an appropriate test statistic for detecting a change in one or a small number of series, while the sum is appropriate for detecting a change in many series \cite[though see][for other approaches to combining statistics across data streams]{enikeeva2019high,fisch2021subset,tickle2021computationally}. 

To investigate the potential for such a method we have run a simulation study comparing an approach which uses both the maximum and sum of the \Focus{} statistics across data streams, and compared to the \texttt{ocd} method of \cite{chen2022high}. We used a similar simulation set-up to \cite{chen2022high}, with data from $100$ data streams. As suggested in \cite{chen2022high}, we used simulation from a model with no change to choose thresholds for both the maximum and sum statistics, and for the three \texttt{ocd} statistics, so that both the \Focus{} method and \texttt{ocd} had an average length of 5,000. We then compared methods based on average detection delay for changes of different size and affecting different numbers of data streams. The results are shown in Table \ref{tab:multiv-sim}.

The \texttt{ocd} method calculates the sequential-Page statistic for a grid of change values for each stream. It then combines these using the maximum. It also uses two statistics based on averaging the log likelihood-ratio test statistic for a change at times corresponding to times of changes with large sequential-Page statistics for individual series. One of these is like our sum statistic, but it assumes the same change time in all series, unlike our implementation which sums test statistics which can correspond to changes at different times. The other statistic for combining across data streams is designed to pick up sparse changes but where multiple streams change.

The results show that if the pre-change mean is known, \texttt{ocd} tends to perform better than our simplistic way of merging the \FocusZ{} statistics. Though our approach is able to detect sparse and large changes more quickly, because the \texttt{ocd} method analyses each data stream using a grid with a relatively small largest change size. However, if the pre-change means are unknown and have to be estimated from training data of length 500, we see that accounting for this by using \Focus{} leads to substantially better performance. The gain in performance will clearly depend on how much training data there is.
}

{ \Focus{} is not an online algorithm, as the number of quadratics to maximise per iteration can fluctuate and is unbounded. We suggested two ways to implement an online version by maximising only at most $P$ quadratics at each iteration based on maximising those that are optimal for a geometrically-space grid of $\mu_1$ values. Such a method will be uniformly better than using sequential-Page with the same grid. However there are others, potentially better, ways of choosing which quadratics to maximise. First one could cycle through quadratics, so that at time $t+1$ you start maximising quadratics that were not considered at time $t$. Alternatively you could use the value of the observation to help choose which quadratics to maximise. For example, if one receives observation $x_t>0$ then this will increase $Q_t(\mu)$ only for $\mu\in(0,2x_t)$, and most at $\mu=x_t$. Thus  there is no point to consider quadratics that are optimal only for values larger than $2x_t$, and one could give priority to quadratics that are optimal for regions closest of $\mu$ closest to $x_t$.
}

In line with Theorem  \ref{th:numberofchanges}, one further possible area of development would be on studying the explicit distribution of the number of quadratics under the alternative, in case the statistics does not reach the threshold. In such case, we expect the number of changes to increase at a faster rate then when under the null: an additional test could be placed on the expected number of quadratics as a fail-safe mechanism to declare a change.

Following the proof of Theorem \ref{th:numberofchanges} it is fairly easy to show that the set of candidate changepoints stored by the offline pDPA algorithm \cite{Rigaill} run for one change is included in the set of changepoints stored by \Focus{}. This provides a bound on the expected complexity of the pDPA for one change. Future work may consider extending the proof of Theorem \ref{th:numberofchanges} to get the expected complexity of pDPA for more than one change or of other functional pruning algorithms such as FPOP \citep{Maidstone} or GFPOP \citep{hocking2020constrained}.

\bibliography{bibliography}

\newpage

\appendix

\begin{center}
{\large\bf SUPPLEMENTARY MATERIAL}
\end{center}

\input{new_appendix}

\end{document}

%% file: sim_study_v3.tex
\subsection{Simulation Study}
\label{sec:simulation-study}

We study the average run length and the detection delay of the \FocusZ{} procedure and its approximation (introduced at the end of Section \ref{sec:focus0_computational}), the sequential Page-CUSUM statistics from equation \eqref{eq:seq_page-gaus}, and the MOSUM procedure from equation \eqref{eq:MOSUM} in case of a pre-change mean known at 0.

The study is structured as follows: 
    For the Page recursion we employ a geometric grid \cite[as recommended by][]{chen2022high}. We use a 10 point grid as that is equivalent to the expected number of intervals stored in \FocusZ{} over a sequence of one hundred thousand observations. To see the potential benefits of using a finer grid, we also use a 20 point grid. We call these two approaches Page-20p, and Page-10p.
    We evaluate MOSUM over a set of 20 window sizes, with these sizes geometrically increasing, and chosen to give MOSUM greatest power for the size of changes specified by the grid used for Page-20p. 
    We also measure performance of the \FocusZ{} approximation on the 10 points grid used for Page-10p. We call this approximation \FocusZ{}-10p.

\begin{figure}[tb]
    \centering
    \includegraphics[width=.45\linewidth]{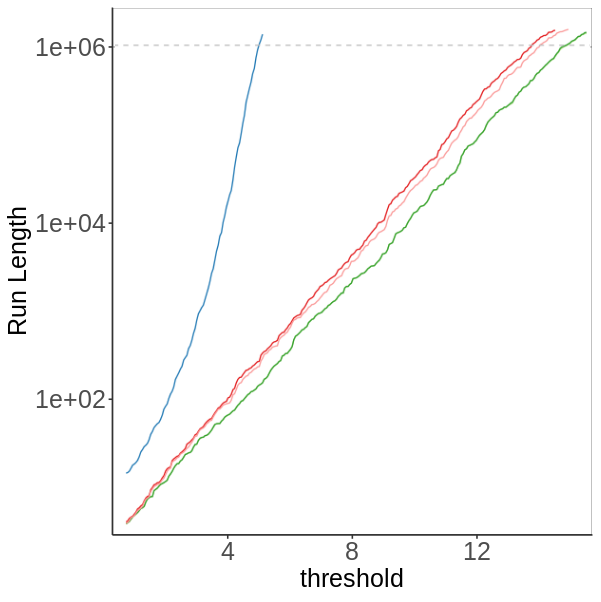}
    \caption[Average Run Length (log scale) in function of the threshold]{Average Run Length against threshold for \FocusZ{} and \FocusZ{}-10p (both identical, and shown in green), Page-20p (pink),  Page-10p (red) and MOSUM (blue). Results are averaged across 100 simulations. We use a log scale on the y axis.}
    \label{fig:avgrun-known}
\end{figure}

For each of these methods, we first estimate the run length as a function of the threshold based on data, with no change, of length two million observations. We average the results across 100 different replicates and summarise them in Figure \ref{fig:avgrun-known}.
For each method we choose the threshold that gives an average run length of $1\e{6}$ observations, and evaluate the detection delay with this threshold over a range of different change magnitudes.
To generate profiles with changes we superimpose on the previous 100 null profiles a piece-wise constant signal with a change at $1\e{5}$. For simplicity we only show changes with a positive increase in magnitude, however the study extends to negative changes too. 

\begin{figure}
    \centering
    \includegraphics[width=\linewidth]{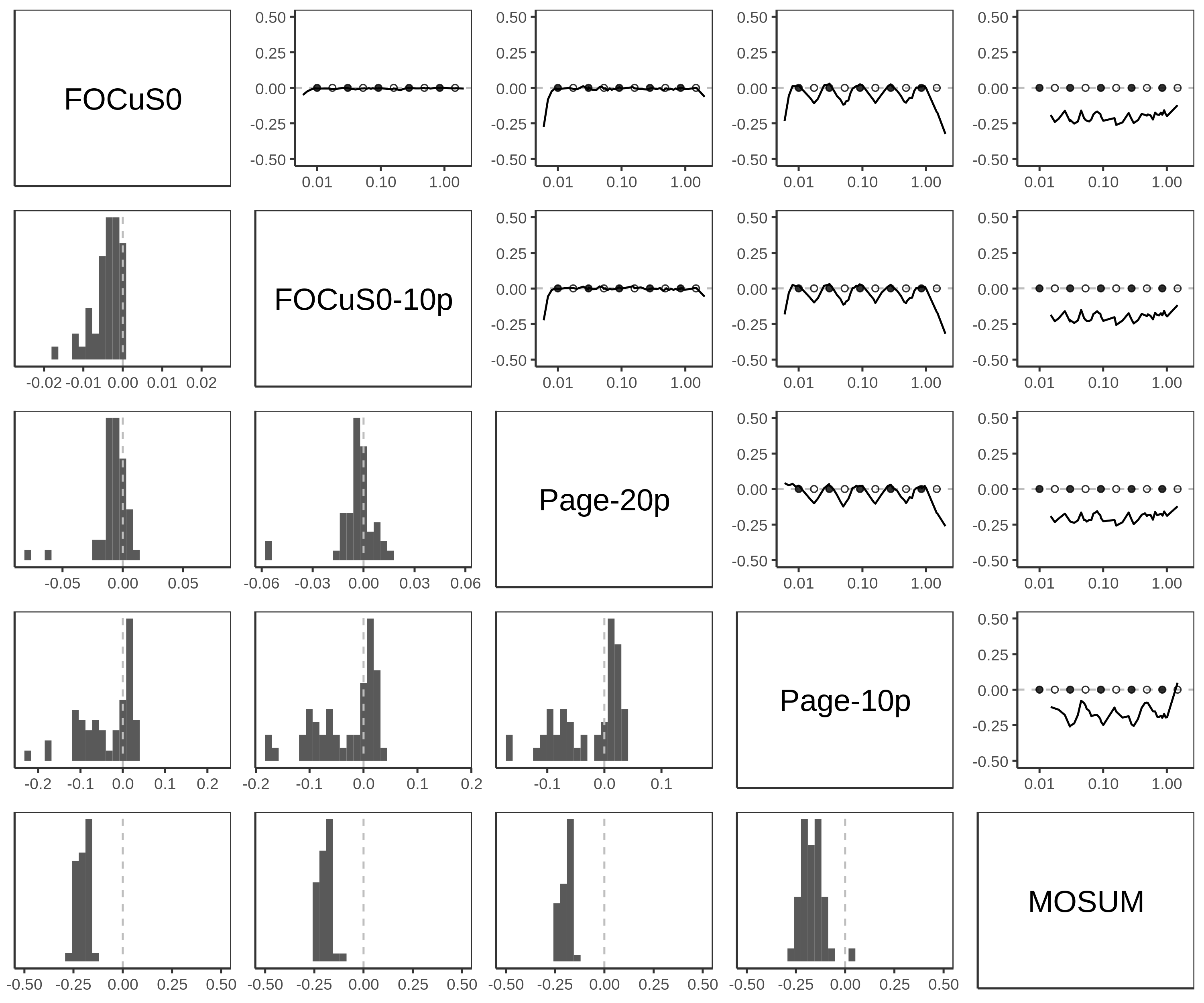}
    \caption[Log ratios of the average detection delays]{
    Plot matrix for log-ratios of the average detection delays of each method considered in the study. The diagonal indicates the name of the i-th tested method.
The (i,j)-th plot shows the log-ratios of the average detection delays of methods i and j, with the highest index method in the numerator and the lower in the denominator, so values below zero mean the lower index method has a lower detection delay. For example, the (1, 2) and (2, 1) plots give the log-ratios between \FocusZ{} against \FocusZ{}-10p, and a value below 0 would mean \FocusZ{} has the smaller average detection delay.
The plots of the upper diagonal show the log-ratio as a function of the change magnitude; the dots indicate the 20 points grid, with the filled ones being the one in common with the 10 points grid. The plots on the lower diagonal show histograms of all the log-ratios, with the vertical dotted line being at $0$.}
    \label{fig:det-delay-known}
\end{figure}

In Figure \ref{fig:det-delay-known} we report log-ratios of the average detection delay for pairs of methods against the different change magnitudes. The most striking comparison is between \FocusZ{} and Page-10p. The relative performance of the methods depends on how the size of change matches with grid used for Page-10p. If these match exactly, then Page-10p has a slightly smaller average detection delay because it has a smaller threshold (see Figure \ref{fig:avgrun-known}). However as we move to changes that are different from the grid points, the Page-10p method loses power relative to \FocusZ{} and the latter can be substantially faster at detecting a change. Once we increase the grid size to 20, we have similar qualitative patterns but now the quantitative differences in performance are small, as is the difference between \FocusZ{} and \FocusZ{}-10p. In general, the biggest gains of \FocusZ{} over the grid-based methods are seen for changes that are smaller than the minimum jump size, or larger than the maximum jump size of the grid. Finally, we see that MOSUM gives noticeably the worst performance of all methods.

%% file: new_appendix.tex
\setcounter{page}{1}
\section{Proof of Proposition \ref{prop1}} \label{app:proof-prop1}

It is straightforward to show, for example by induction, that the solution $Q_n(\mu)$ to recursion (\ref{eq:functional-recursion}) can be written in the form
\[
Q_{n}(\mu)=\max_{s=0,\ldots,n} \left\{\sum_{t=s+1}^{n} \mu \left(x_t-\frac{\mu}{2}\right) \right\},
\]
where we treat the sum from $n+1$ to $n$ as empty, and hence equal to 0.
Hence
\begin{eqnarray*}
\max_\mu Q_{n}(\mu) & = & \max_{s=0,\ldots,n} \left\{ \max_{\mu}\sum_{t=s+1}^n \mu \left(x_t-\frac{\mu}{2}\right) \right\} \\
& = & \max_{s=0,\ldots,n-1} \left\{
\frac{1}{2} \sum_{t=s+1}^n \left(
\frac{\sum_{j=s+1}^n x_j}{n-s}
\right)^2
\right\} \\
&=& \max_{w=1,\ldots,n} \left\{
\frac{1}{2} w \left(
\frac{\sum_{j=n-w+1}^n x_j}{w}
\right)^2
\right\}
\end{eqnarray*}
The second line uses the fact that the maximum over $\mu$ is when $\mu$ is the sample mean of $x_{\tau:n}$. For the second step we use the fact that the maximum is never the empty sum, and thus we can drop the case $s=n$ from the maximisation. The terms in the final expression are just $(1/2) M_w(n)^2$ as required. The result in terms of $P(n)$ follows directly from $P(n)=\max_w M_w(n)$. \hfill $\Box$

\section{On the expected number of changes stored by \Focus{}}\label{app:number_of_ints}

\subsection{Variants of the FOCuS implementations}

We study the number of candidate changepoints $\tau \in \{1, \cdots, n\}$ stored by FOCuS at each iteration. We report the  possible FOCuS optimizations introduced in the main body of the paper:

\begin{itemize}
    \item \FocusZ{} which solves the problem for a known pre-change mean $\mu_0$ (typically 0) and unknown post-change mean $\mu_1$.
\begin{equation}
    \mathcal{Q}^{0}_n = \underset{\substack{\tau \in \{ 1, \dots, n \} \\ \mu_0=0, \mu_1 \ \in \ \mathbb{R}} }{\max} -\left\{\sum_{t=1}^{\tau} (x_t - \mu_0)^2 - \sum_{t=\tau+1}^n (x_t - \mu_1)^2 \right\}.
\end{equation}
   This problem is solved through Algorithm \ref{alg:melkmans}, an algorithm similar to the Melkman's algorithm \cite[]{melkman1987line}.
    \item \Focus{} which solves the problem for both unknown pre-change and post-change means: 
\begin{equation}
    \mathcal{Q}_n = \underset{\substack{\tau \in \{ 1, \dots, n \} \\ \mu_1, \mu_0 \in \mathbb{R}} }{\max} \left\{ -\sum_{t=1}^{\tau} (x_t - \mu_0)^2 - \sum_{t=\tau+1}^n (x_t - \mu_1)^2 \right\}.
\end{equation}

\end{itemize}

\subsection{Assumptions and definitions}

In the rest of this section, we let $x_1, \dots x_n$ be our ordered sequence of observations. To denote a subset of such sequence, we will write, from $i$ to $j$: $x_{i:j} = x_i, \dots, x_j$ for $i < j$. We denote a true changepoint with $\tau^*$.

We assume that:
\begin{equation}\label{eq:data}
x_i = \mu_i + \varepsilon_i,
\end{equation}
where $\varepsilon_i$ are i.i.d with a continuous distribution and $\mu_i$ is a piecewise constant signal in $1$ or $2$ pieces. 

We define the cost of a segmentation $x_{i:j}$ with a change at $\tau$ as:
\begin{displaymath}\label{eq:onechange}
q_{i:j, \tau}(\mu_0, \mu_1) = \sum_{t=i}^{\tau} (x_t - \mu_0)^2 + \sum_{t=\tau+1}^j (x_t - \mu_1)^2.
\end{displaymath}
with pre-change and post-change means $\mu_0$ and $\mu_1$. As a convention, for $j = \tau$, we take: $q_{i:j, j}(\mu_0, \mu_1) = \sum_{t=i}^j (x_t - \mu_0)^2$.

\paragraph{Sets of candidate changepoints} We call  $\mathcal{I}^{0}_{i:j}$ the set of candidate changepoints stored by $\text{FOCuS}^0$ for $\mu_1>0$, and $\mathcal{I}_{i:j}$ the set stored by FOCuS for $\mu_1>\mu_0$. 
By definition those will be:

\begin{eqnarray*}
\mathcal{I}^{0}_{i:j}  & =  & \left\{ \tau \ | \ \exists \ \mu_1 > 0, \ \forall \tau' \neq \tau, \qquad q_{i:j, \tau}(0, \mu_1) \ < \qquad q_{i:j, \tau'}(0, \mu_1) \right\}, \\
\mathcal{I}_{i:j} & =  & \left\{ \tau \ | \ \exists \ \mu_1 > \mu_0, \forall \tau' \neq \tau, \qquad q_{i:j, \tau}(\mu_0, \mu_1) <  \qquad q_{i:j, \tau'}(\mu_0, \mu_1) \right\},
\end{eqnarray*}

By definition $\mathcal{I}^{0}_{i:j}$ is in $\mathcal{I}_{i:j}$
Our goal will then be to control the size of the set $\mathcal{I}_{i:j}$ of candidate changepoints stored by \Focus{}.




\subsection{Main results}

On the assumption of a realization from \eqref{eq:data} we can get the following bound on the number of changepoints stored by \FocusZ{} and \Focus{}. 

\begin{theorem}\label{th:numberofchanges}
For all $n\geq 1$
\begin{equation*}
    E(\#\mathcal{I}^{0}_{1:n}) \leq E(\#\mathcal{I}_{1:n}). 
\end{equation*}

If { $\mu_i$ is constant with respect to $i$}:
\begin{equation*}
    E(\#\mathcal{I}_{1:n}) = 1+ \sum_1^{n-1} 1/(t+1) \leq (1 + \log(n))
\end{equation*}
and
%
{ if $\mu_i$ has a single changepoint}, we have
\begin{equation*}
    E(\#\mathcal{I}_{1:n}) \leq 2(1 + \log(n/2)).
\end{equation*}
\end{theorem}

By symmetry, the same result holds for the number of quadratics stored by \FocusZ{} for $\mu_1<0$ and \Focus{} for $\mu_1 < \mu_0.$

\paragraph{Overview of the proof} The proof to this theorem relies on a combination of three lemmas, summarized here:
\begin{enumerate}
    \item Lemma \ref{lem:inclusion2}: for $i \leq j < k$ we have $\mathcal{I}_{i:k} \subseteq \mathcal{I}_{i:j} \cup \mathcal{I}_{j+1:k}$; 
    \item Lemma \ref{lem:convexhull}: $\mathcal{I}_{i:j}$ are the extreme points of the largest convex minorant of the sequence $S_{i:j}$, where ~{ $S_{t}=\sum_{k=1}^t x_k$}; 
    \item Lemma \ref{lemma:Andersen}: that controls the number of extreme point of a random-walk \citep[derived from][]{andersen1955fluctuations, abramson2012some}. 
\end{enumerate}
The three lemmas are covered in details and proven in Appendix \ref{app:FOCuSlemmas}.\\

{\bf{Proof}}:
By definition $\mathcal{I}_{i:j}$ includes $\mathcal{I}^{0}_{i:j}$ and we get the first inequality.
For { the case where $\mu_i$ is constant with respect to $i$} we apply Lemma \ref{lemma:Andersen}. 
For { the case where $\mu_i$ has a single changepoint}, using Lemma \ref{lem:inclusion2} we get that $$\mathcal{I}_{1:n} \subseteq 
\mathcal{I}_{1:\tau^*} \cup \mathcal{I}_{\tau^*+1:n}.$$
We then apply Lemma \ref{lemma:Andersen} on $\mathcal{I}_{1:\tau^*}$ and $\mathcal{I}_{\tau^*+1:n}.$ The worst case is obtained for $\tau^* = n/2$. 
\hfill $\Box$

An empirical evaluation of this bound can be found in Appendix \ref{sec:FOCuSTheoEMPIRICAL}.

\subsection{Inclusion and convex hull Lemmas}\label{app:FOCuSlemmas}






\paragraph{A useful identity}
For any { $i\leq\tau<\tau'\leq j$}, $\mu_0$ and $\mu_1$ we have that
\begin{equation}\label{eq:noIJ}
q_{i:j, \tau}(\mu_0, \mu_1) - q_{i:j, \tau'}(\mu_0, \mu_1)  =  (\mu_0 - \mu_1) \left( 2 \sum_{\tau+1}^{\tau'} x_t - (\tau'-\tau)(\mu_0 + \mu_1)\right), 
\end{equation}
which does not depend on $i$ and $j$. This identity simplifies the proof of the following lemmas.

\begin{lemma}\label{lem:inclusion2}
For $i \leq j \leq k$
\begin{equation}
\mathcal{I}_{i:k} \quad \subseteq \quad \mathcal{I}_{i:j}  \cup \mathcal{I}_{j+1:k}
\end{equation}
\end{lemma}

{\bf Proof}: 
Consider any $\tau$ in $(i+1:j) \cap \mathcal{I}_{i:k}$, by definition 
\begin{equation*}
\exists \ \mu_0 < \mu_1, \forall \ \tau' \neq \tau \ with \ \tau' \ in \  (i+1:k), \quad q_{i:k, \tau}(\mu_0, \mu_1) < q_{i:k, \tau'}(\mu_0, \mu_1),
\end{equation*}
then using equation \eqref{eq:noIJ} we get $q_{i:j, \tau}(\mu_0, \mu_1) < q_{i:j, \tau'}(\mu_0, \mu_1)$ and thus $\tau$ is also in $\mathcal{I}_{i:j}.$
We proceed similarly for any $\tau$ in $(j+1:k) \cap \mathcal{I}_{i:k}$. $\hfill \square$


The next lemma relates the set $\mathcal{I}_{i:j}$ to the lower convex hull of $S_{i:j}$.
\begin{lemma}\label{lem:convexhull}
The set of $\tau$ in $\mathcal{I}_{i:j}$ are the extreme points of the largest convex minorant of the sequence $S_{i:j}$. 
\end{lemma}

{\bf Proof}:
 
Assume that $\tau'$ is in $\mathcal{I}_{i:j}$. Then, by definition of $\mathcal{I}_{i:j}$, there exists a $(\mu_0,\mu_1)$ with $\mu_1>\mu_0$
such that for all $\tau<\tau'$,
\[
q_{i:j,\tau}(\mu_0,\mu_1)-q_{i:j,\tau'}(\mu_0,\mu_1) > 0
\]
and for all $\tau''>\tau'$
\[
q_{i:j,\tau'}(\mu_0,\mu_1)-q_{i:j,\tau''}(\mu_0,\mu_1)<0.
\]
Using  equation \eqref{eq:noIJ}, then the equation for $\tau<\tau'$ gives
\begin{eqnarray*}
(\mu_0 - \mu_1) \left( 2 \sum_{\tau+1}^{\tau'} x_t - (\tau'-\tau)(\mu_0 + \mu_1)\right) & > & 0 \\
\Rightarrow \left( 2 \sum_{\tau+1}^{\tau'} x_t - (\tau'-\tau)(\mu_0 + \mu_1)\right) & < & 0 \\
\Rightarrow \bar{x}_{\tau+1:\tau'} < \frac{1}{2}(\mu_0+\mu_1),
\end{eqnarray*}
where $\bar{x}_{\tau+1:\tau'} = \frac{\sum_{t=\tau+1}^{\tau'} x_t}{\tau'-\tau}.$ The second inequality follows from $\mu_0-\mu_1<0$.

A similar argument for $\tau''>\tau'$ gives that 
\[
\bar{x}_{\tau'+1:\tau''} > \frac{1}{2}(\mu_0+\mu_1).
\]
For such a $(\mu_0,\mu_1)$  to exist such that these inequalities hold for all $\tau<\tau'<\tau''$ we need that 
$$\bar{x}_{\tau+1:\tau'} < \bar{x}_{\tau'+1:\tau''},$$
for all $\tau<\tau'<\tau''$.

Re-writing the sample means in terms of the cumulative sums of the data, this is equivalent to for all  $\tau$ and $\tau''$ satisfying $\tau<\tau'<\tau''$: 
$$\frac{S_{\tau'} - S_{\tau}}{\tau'-\tau} < \frac{S_{\tau''} - S_{\tau'}}{\tau''-\tau'},$$
and therefore $(S_{\tau'},\tau')$ is part of the largest convex minorant of the sequence $S_{i:j}.$

 We now prove the converse. Assume that $\tau'$ is part of the convex minorant of $S_{i:j}.$ Then for all $\tau$ and $\tau''$ satisfying $\tau<\tau'<\tau''$: 
$$\frac{S_{\tau'} - S_{\tau}}{\tau'-\tau} < \frac{S_{\tau''} - S_{\tau'}}{\tau''-\tau'},$$ and therefore there exists a $m$ such that
$$\max_{\tau} \bar{x}_{\tau+1:\tau'} < m <\min_{\tau''} \bar{x}_{\tau'+1:\tau''}.$$

Now picking $\mu_0$ and $\mu_1$ such that $\frac{1}{2}(\mu_0+\mu_1)=m$ and  $\mu_0 - \mu_1= - 1$ we get using equation \eqref{eq:noIJ} we get
 that for all $\tau<\tau'$,
\[
q_{i:j,\tau}(\mu_0,\mu_1)-q_{i:j,\tau'}(\mu_0,\mu_1) > 0
\]
and for all $\tau''>\tau'$
\[
q_{i:j,\tau'}(\mu_0,\mu_1)-q_{i:j,\tau''}(\mu_0,\mu_1)<0.
\]


$\hfill \square$



The next lemma is based on \cite{andersen1955fluctuations}.
\begin{lemma}\label{lemma:Andersen}
Assuming the $x_t$ follow an i.i.d continuous distribution on $i:j$ 
then 
$E(\#\mathcal{I}_{i:j})=\sum_{t=1}^{j-i-1} 1/(t+1) +1 \leq \log(n) + 1$.
\end{lemma}

{\bf Proof}:
We use Lemma \ref{lem:convexhull} and then apply \cite{andersen1955fluctuations} (definitions at pages 195 and 196, then result of page 217.   \cite{andersen1955fluctuations} does not include the vertex at the final point of the random walk, so we need to add a $1$ for the point at $j$. Standard results for the harmonic series show that $\sum_{t=1}^{j-i-1} 1/(t+1) \leq \log(j-i)$. The proof follows as $(j-i)\leq n$.
$\hfill \square$

\section{Additional Empirical Results}

\subsection{Comparison to Lorden (1971)}\label{App:Lorden}

\begin{figure}
    \centering
    \includegraphics[scale=0.4]{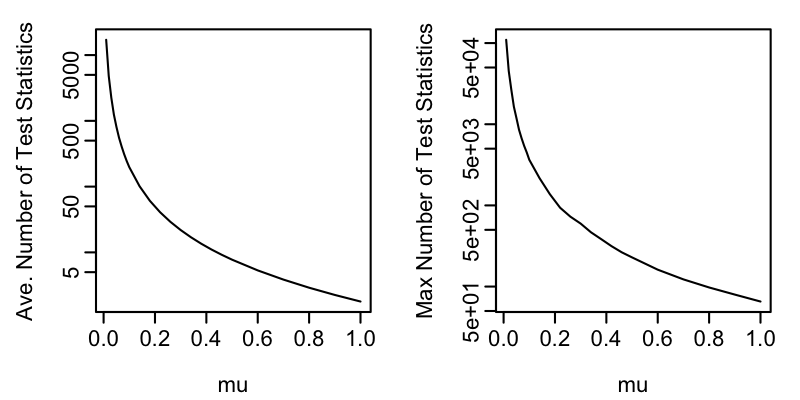}
    \caption{Plot of expected and maximum number of test statistics stored/evaluated by Lorden (1971) for data of length $10^6$, against the minimum change size (mu). The $y$-axes are on a log-scale.}
    \label{fig:Lorden1}
\end{figure}

We implemented Lorden (1971)'s algorithm for detecting a positive change. This algorithm requires the specification of a minimum change size, $\mu^*$ say. It then runs the sequential-Page procedure for $\mu^*$. This procedure will reset, i.e. the sequential-Page statistic will be 0, at certain times. At these times, the sequential-Page statistic calculated for any change $\mu>\mu^*$ would be 0. Thus to detect changes bigger than $\mu^*$ Lorden (1971) proposes calculating the likelihood ratio test statistic for a change starting at any time since the last reset. I.e., if at time $t$ the last reset was at time time $\tau$, then it would calculate the likelihood ratio statistic for a change at each of times $\tau,\tau+1,\ldots,t-1$. The final test statistic is the maximum over these. This involves calculating $t-\tau$ statistics, and is equivalent in cost to maximising $t-\tau$ quadratics in \FocusZ{}.

\begin{figure}
    \centering
    \includegraphics[scale=0.8]{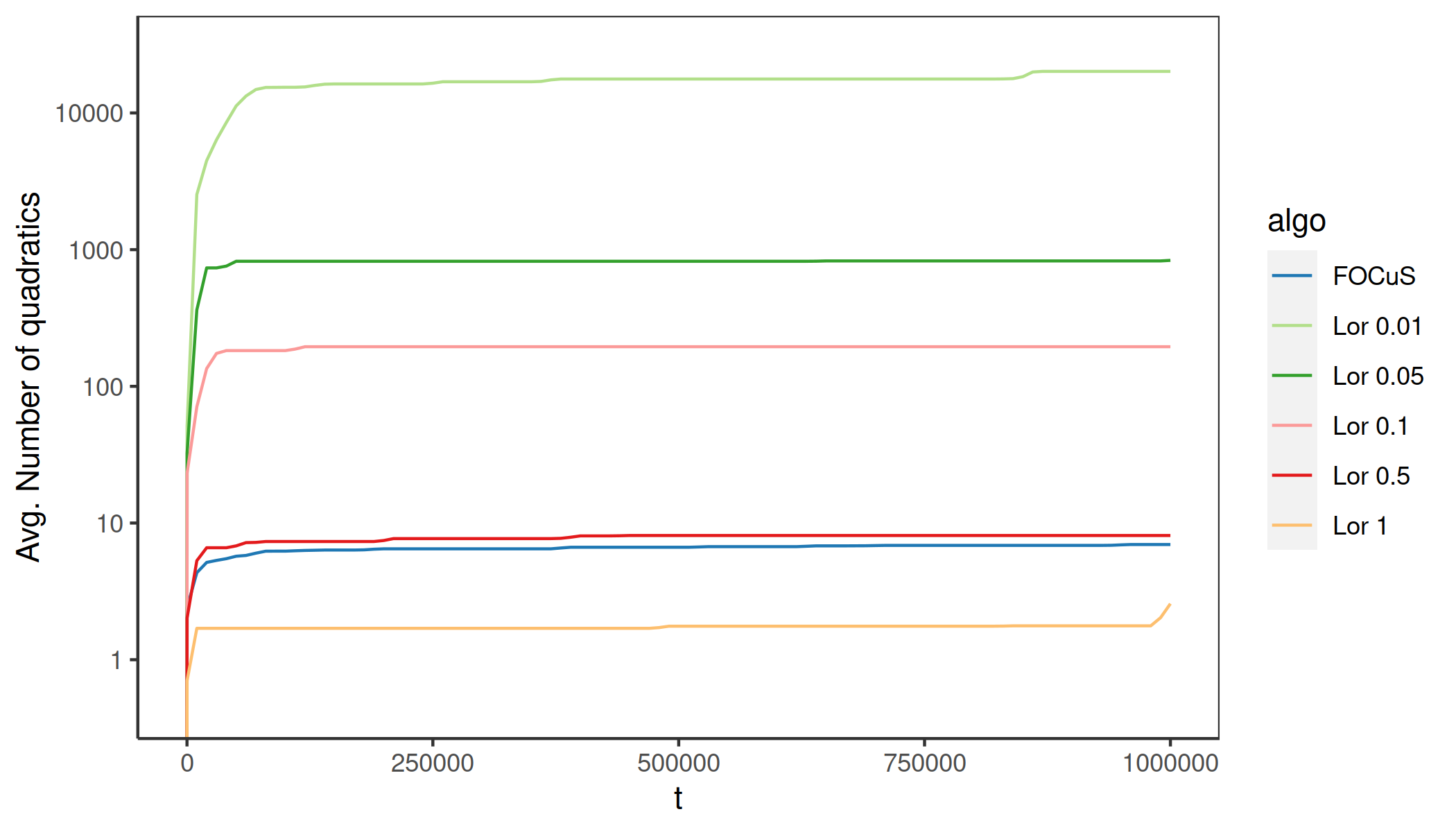}
    \caption{Plot of expected number of test statistics stored/evaluated by Lorden (1971) and expected number of quadratics stored by \FocusZ{} against time for data of length $10^6$. Results obtained by fitting a monotone mean function to data from 50 replication of each algorithm. The $y$-axis is on a log-scale.}
    \label{fig:Lorden2}
\end{figure}

Figure \ref{fig:Lorden1} shows the average and maximum number of test-statistics stored by Lorden (1971)'s algorithm as a function of the minimum jump size. These have been calculated empirically for data of length $n=10^6$, with the results averaged across 50 replications. We also show the expected number of statistics stored as a function of $t$ for different minimum change sizes, and the expected number of quadratics stored by \FocusZ{} in Figure \ref{fig:Lorden2}

For data of length $10^6$, \FocusZ{} would store around 8 quadratics on average. Thus to be computationally more efficient, Lorden (1971) would require a minimum change size of around 0.5 or higher. Furthermore, the variability in the number of test statistics stored by Lorden (1971) is much higher than for \Focus{} (e.g. for a minimum change size of 0.8 it can require evaluating over 100 test statistics) which can be problematic if there is a limit on the computational resource available per iteration.

\subsection{Empirical bound evaluation}\label{sec:FOCuSTheoEMPIRICAL}
\begin{figure}[!htb]
    \centering
    \includegraphics[height=.45\linewidth, clip=true, trim=0cm 0cm 3cm 0cm]{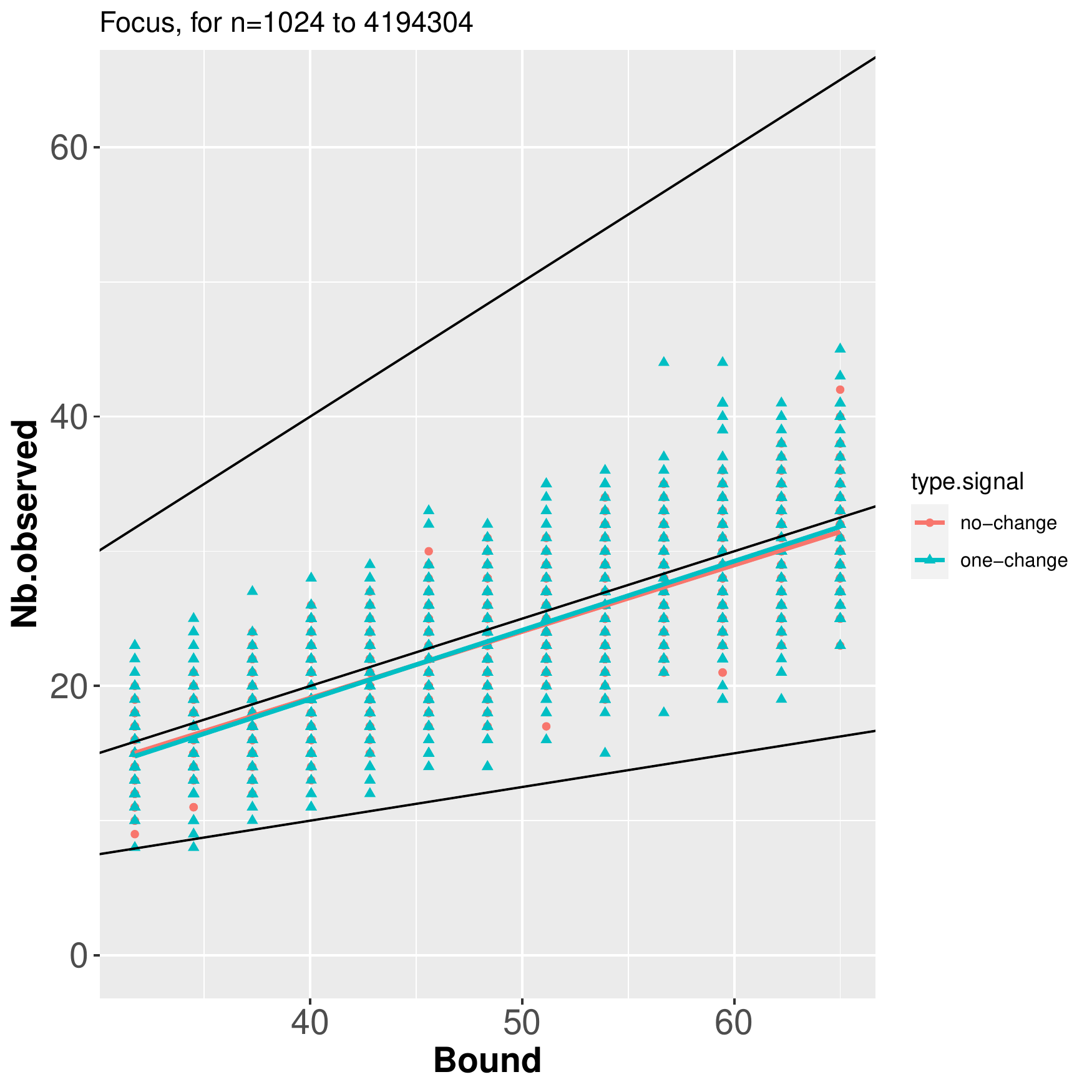}
      \includegraphics[height=.45\linewidth, clip=true, trim=0cm 0cm 0cm 0cm]{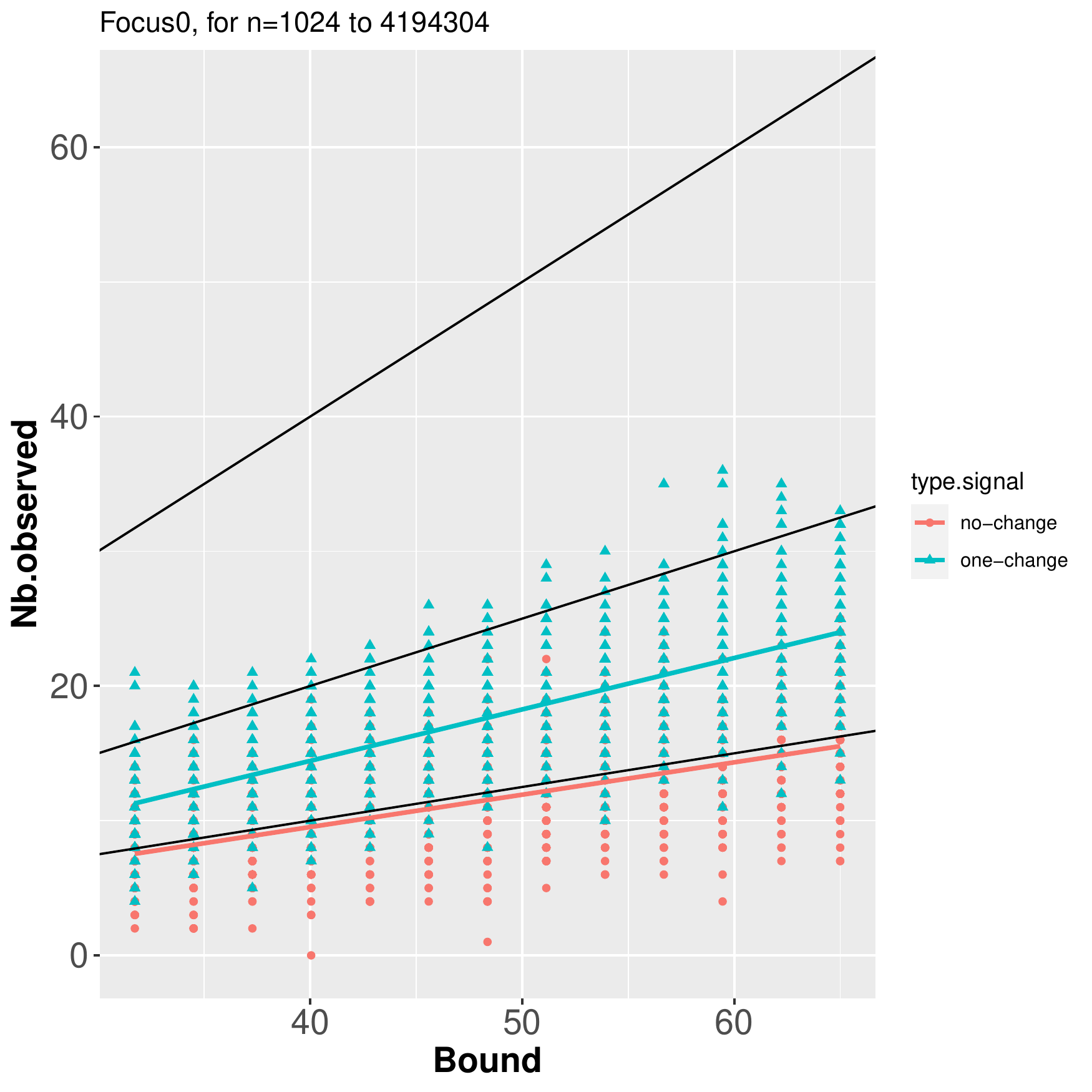}
    \caption{Number of observed quadratics stored by \Focus{} (left) and \FocusZ{} (right) for signals with no change or with one change  against the bound $4(\log(n)+1)$. The three black lines represent the function $y=x$, $y=0.5x$ and $y=0.25x$, representing the lines $4(\log(n)+1)$, $2(\log(n)+1)$ and $\log(n)+1$ respectively. Red and blue line are the fitted regression lines for respectively the no-change and single-change scenarios.}
    \label{fig:numberofcandidates}
\end{figure}

To illustrate the bound of Theorem \ref{th:numberofchanges} we simulate signals of various length $n$ (from $n=2^{10}$ to $n=2^{22}$) without change (100 replicates) and with one change (100 replicates). We then record the number of candidates stored by \Focus{} (left) and \FocusZ{} (right). Where a change was present, its location was sampled uniformly between 1 and $n-1$. Similarly, the change magnitude was sampled uniformly at random in $[0, 4].$ Results are summarised in Figure \ref{fig:numberofcandidates}. 

 The bound in Theorem \ref{th:numberofchanges} applies to both \FocusZ{} and \Focus{} for an up change, and a similar bound holds by symmetry for down changes. These give a total bound on the number of quadratics stored that is $2(\log(n)+1)$ if the data is simulated without a change and $4(\log(n/2)+1)$ if the data is simulated with a single change. We plot the observed number of quadratics stored against $4(\log(n)+1)$, and also plot the lines $2(\log(n)+1)$ and $\log(n)+1$. For \Focus{} we see that the bound of $2(\log(n)+1)$ for the case where the data is simulated without a change is tight. However the number of quadratics stored is very similar for data simulated with a change -- suggesting that the $4(\log(n/2)+1)$ bound for that case is conservative. For \FocusZ{} the number of quadratics stored is empirically half that stored by \Focus{} for the case where the data is simulated without a change, in line with the discussion after Theorem \ref{th:numberofchanges}. The bound in the theorem is based on a bound on the number of vertices on the convex minorant of the random walk defined by the cumulative sum of the data. However, for \FocusZ{} the number of quadratics stored is equal to a subset of the vertices, namely the vertices that are followed by a side of the minorant with positive slope. By symmetry, this would be the vertex at the minimum of the convex minorant plus, in expectation, half the remaining vertices.

\section{Estimation of initial parameters}\label{sec:param-est}

We propose a simple sequential for the initial parameters needed to run \RFocus{} (and the other implementations) in an semi-supervised manner.  The general idea is to first fine the value the $K$ parameter of the bi-weight loss, and based on that tune the $\lambda$ threshold value on a probation period. Say for each sequence we consider the first $w$ observations for training. For tuning the $K$ parameter of the bi-weight loss, one could simply store the initial $w$ values and, if any observation within $1.5$ interquantile ranges of those values are present, \textit{i.e.} if we are in presence of outliers, then simply pick the $K$ highest occurring value within this limit.
On the same period we estimate the variance $\sigma^2$ and run the \RFocus{} procedure on the normalised values of the probation period with the value $K$, recording the trace of the statistics at each iteration $\mathcal{Q}_1, \dots, \mathcal{Q}_w$. We then set $\lambda = \kappa \times \max_{i = 1, \dots, w} \mathcal{Q}_i$ for some $\kappa \in \mathbb{R}^+$. In the presented application we set $\kappa = 1.5$ at the beginning and to $\kappa = \log(\tau_s) / \log(\tau_s - \tau_{s-1})$ when restarting the algorithm after a detection, with $\tau_s, \ \tau_{s-1}$ being the last and previous last stopping times.

In total, this estimation adds a linear in $w$ computational overhead, which consist in temporarily storing the initial $w$ values and performing the estimation of the $K$.